\documentclass[fleqn,usenatbib,usedcolumn]{mnras}
     \usepackage[british]{babel}             
    \usepackage{newtxtext}                  
   \let\la=\lesssim 
  \usepackage[T1]{fontenc}                
  \usepackage{graphicx}                   
 \usepackage{subfig}

\usepackage{amsmath}	
\usepackage{amssymb}	

\title[Imaging of SNR IC443 and W44 with SRT]{Imaging of SNR IC443 and W44 with the Sardinia Radio Telescope at 1.5 GHz and 7 GHz}

\author[]{
E. Egron$^{1}$,
A. Pellizzoni$^{1}$,
M.\,N. Iacolina$^{1}$,
S. Loru$^{1,2}$,
M. Marongiu$^{1}$,
S. Righini$^{3}$,
\newauthor 
M. Cardillo$^{4}$,
A. Giuliani$^{5}$,
S. Mulas$^{2}$,
G. Murtas$^{2}$,
D. Simeone$^{2}$,
R. Concu$^{1}$,
\newauthor A. Melis$^{1}$,
A. Trois$^{1}$,
M. Pilia$^{1}$, 
A. Navarrini$^{1}$, 
V. Vacca$^{1}$,
R. Ricci$^{3}$,
G. Serra$^{1}$,
\newauthor 
M. Bachetti$^{1}$, 
M. Buttu$^{1}$,
D. Perrodin$^{1}$,
F. Buffa$^{1}$,
G.\,L. Deiana$^{1}$,
F. Gaudiomonte$^{1}$,
\newauthor
A. Fara$^{1}$,
A. Ladu$^{1}$,
F. Loi$^{1,2}$,
P. Marongiu$^{1}$,
C. Migoni$^{1}$,
T. Pisanu$^{1}$,
S. Poppi$^{1}$,
\newauthor
A. Saba$^{1}$,
E. Urru$^{1}$,
G. Valente$^{1}$,
G.P. Vargiu$^{1}$
\\
$^{1}$INAF, Osservatorio Astronomico di Cagliari, Via della Scienza 5,
09047 Selargius, Italy\\
$^{2}$Dipartimento di Fisica, Universit\`a degli Studi di Cagliari, SP
Monserrato-Sestu, KM 0.7, 09042 Monserrato, Italy\\
$^{3}$INAF, Istituto di Radio Astronomia di Bologna, Via P. Gobetti 101, 40129 Bologna, Italy\\
$^{4}$INAF, Osservatorio Astrofisico di Arcetri, Largo E. Fermi 5, 50125 Firenze, Italy\\
$^{5}$INAF, Istituto di Astrofisica Spaziale e Fisica cosmica di Milano, via E. Bassini 15, 20133 Milano, Italy
}

\date{Accepted XXX. Received YYY; in original form ZZZ}

\pubyear{2016}

\begin{document}
\label{firstpage}
\pagerange{\pageref{firstpage}--\pageref{lastpage}}
\maketitle

\begin{abstract}
Observations of supernova remnants (SNRs) are a powerful tool for investigating the later stages of stellar evolution, the properties
of the ambient interstellar medium, and the physics of particle
acceleration and shocks.
For a fraction of SNRs, multi-wavelength coverage from radio to ultra-high-energies has been
provided, constraining their contributions to the production
of Galactic cosmic rays.
Although radio emission is the most
common identifier of SNRs and a prime probe for refining models,
high-resolution images at frequencies above 5 GHz are surprisingly
lacking, even for bright and well-known SNRs such as IC443 and W44.
In the frameworks of the Astronomical Validation and Early Science Program
with the 64-m single-dish Sardinia Radio Telescope, we provided, for
the first time, single-dish deep imaging at 7 GHz of the IC443 and W44
complexes coupled with spatially-resolved spectra in the 
$1.5-7$ GHz
frequency range. 
Our images were obtained through on-the-fly mapping techniques, providing antenna beam oversampling and
resulting in accurate continuum flux density measurements.
The integrated flux densities associated with IC443 are
$S_{\mathrm{1.5GHz}}=134\,\pm\,4$ Jy and
$S_{\mathrm{7GHz}}=67\,\pm\,3$ Jy.
For W44, we measured total flux densities of $S_{\mathrm{1.5GHz}}=214\,\pm\,6$ Jy and $S_{\mathrm{7GHz}}=94\,\pm\,4\,\mathrm{Jy}$.
Spectral index maps provide evidence of a wide physical parameter scatter among different SNR
regions: a flat spectrum is observed from the brightest SNR regions
at the shock, while steeper spectral indices (up to $\sim\,0.7$) are
observed in fainter cooling regions, disentangling in this way different
populations and spectra of radio/gamma-ray-emitting electrons in
these SNRs.
\end{abstract}




\section{Introduction}

Supernova remnants (SNRs) result from stellar explosions, typically releasing energy 
in excess of 10$^{50}$ erg in the interstellar medium. The strong  shock waves
associated with SNRs (with typical initial speeds $>$1,000 km/s) heat the
ejecta and ambient gas to X-ray-emitting temperatures until speeds fall below $\sim$100 km/s over a timescale of a few 10,000 years.
These shocks also put a large fraction of their energy into magnetic fields and
accelerated non-thermal radio/gamma-ray-emitting electrons and gamma-ray-emitting heavier
particles, with an important influence on the remnant physics and on
the population of Galactic cosmic-ray particles.
Observations of SNRs are thus important not only for the astronomical understanding
of the intrinsic phenomenology of stellar evolution but also for investigating global
processes in our Galaxy, -e.g.- the properties of the shocked ambient interstellar
medium and the origin of cosmic rays \citep{Ginzburg_64}.
In fact, Supernovae are believed to be the main sources of Galactic cosmic rays
under $\sim$10$^{15}$--10$^{16}$ eV \citep[e.g.,][for a review]{Amato_14,Blasi_13,Reynolds_08}.

The first radio surveys at low frequencies revealed bright and large SNRs
\citep[e.g.][]{Clark_76} displaying non-thermal spectra (power-law flux
$S_{\nu} \propto \nu^{-\alpha}$ with $\alpha \simeq 0.5$). Subsequent sensitive
surveys with higher resolution also identified a number of faint and
compact SNRs.
To date, $\sim$300 Galactic SNRs have been catalogued \citep{Green_14} and well characterised
in the radio band, typically up to 1 GHz. Typical SNR sizes range from a few
tens of square arcminutes to a few square degrees with total fluxes of $1-300$ Jy at $\sim$1 GHz . 
For a fraction of them, multi-wavelength coverage ranging from radio to ultra-high-energies 
(TeV) has been provided, although radio emission is the most
common identifier of SNRs (extended radio sources with steep spectra).
In particular, about 20 SNRs have been firmly detected in gamma-rays mostly through
the observations of Fermi-LAT \citep{Acero_16} and AGILE satellites in the GeV range \citep{Giuliani_11A}
and Cherenkov telescopes in the TeV range \citep{Humensky_15,Hess_11}.


Despite the growing interest in high-energy GeV$-$TeV emission and its connection
with radio emission, multi-wavelength data of SNRs are sparse and often missing.
Synchrotron radio emission is expected at least up to $20-50$ GHz, and is produced
by electrons with energies in the GeV range and magnetic fields of $\sim 10-100~\mu$G.
However, for the most interesting and bright objects, high-resolution images at frequencies $>$5 GHz in the confused regions of the Galactic
Plane are lacking and not easily achievable through interferometry due to the very large
SNR structures.
Recent observations were performed between 10 and 20 GHz
with QUIJOTE \citep{Genova_16}, providing intensity and
polarization maps along the Galactic Plane ($24^{\circ} < l <
45^{\circ}$, $|b|$ $< 8^{\circ}$), with an angular resolution of $\sim
1^{\circ}$. Moreover, Planck observations using the Low Frequency Instrument were
performed between 30 and 70 GHz \citep[][and
references therein]{Onic_15}, but here
again, the resolution cannot provide accurate maps of SNRs at these
frequencies.
Dedicated higher resolution observations are thus required to firmly address and disentangle models.

Integrated fluxes are typically
available in the literature for up to $5-10$ GHz, while SNR spatially-resolved fluxes are largely
unexplored in this frequency range and above.
In particular, a better characterisation of radio spectra (synchrotron
spectral indices and breaks) is necessary for constraining the high-energy spectra in the frame
of IC/bremmstrahlung leptonic models and probe them versus hadronic models. 
For example, the gamma-ray fluxes and spectra are strongly dependent on the IC photon target
parameters and spectral breaks in the electron distribution, and in most cases there are
no unambiguous model solutions without detailed measurements of the latter, since the former is intrinsically uncertain \citep{Cardillo_14,Ackermann_13,Giuliani_11B}.
Furthermore, a careful verification (through single-dish imaging, possibly in combination with interferometric data) of co-spatiality of the
radio high-frequency and gamma-ray emissions could provide a crucial test of the hypothesis of hadron vs. electron emission. 
In fact, recent constraints on cosmic ray emission from SNRs and
related models \citep[e.g.][]{Ackermann_13}
are based on integrated radio fluxes only (no
spatially resolved spectra), implying the simplistic assumption of a single
electron population for the whole SNR.

SNR IC443 and W44 represent ideal targets for better testing the above models,
thanks to their interesting complex morphology and availability of extensive
multi-wavelength data, from radio to gamma rays. 
They belong to the remnant "mixed morphology"
class \citep{Rho_98}, and are characterised by a highly filamentary radio shell (synchrotron emission) and a central
thermal X-ray emission.
IC443 (also named 3C157 and nicknamed the "Jellyfish Nebula") is one
of the best-studied Galactic SNRs \citep{Chevalier_99,Petre_88}, with an estimated age of $\sim 30,000$ yrs derived from the associated neutron star proper motion \citep{Olbert_01},  then revised to $\sim 4,000$ yrs on the basis of X-ray plasma properties \citep{Troja_08}.
Located at about 1.5 kpc in the direction of the Galactic anticenter, the large structure of the source
extends over $0.75^{\circ}$ and shows evidence of interactions with
both atomic and molecular clouds \citep{Lee_08,Snell_05,Cesarsky_99,Dickman_92}. 
Its radio flux density at 1 GHz is about 160 Jy, with a spectral index
$\alpha=0.39\,\pm\,0.01$ in the frequency range 10 MHz $-$ 10.7 GHz \citep{Castelletti_11}.
W44 is a bright radio SNR (S$_{\rm 1GHz} \sim 230$ Jy, spectral index $\alpha=0.37\,\pm\,0.02$ in the  22 MHz $-$ 10.7 GHz band) located in the Galactic Plane at
a distance of $\sim2.9$ kpc \citep[][and references therein]{Castelletti_07}. Formed nearly 20,000 yrs ago \citep{Wolszczan_91,Smith_85}, 
it presents an asymmetric morphology of about half a degree in size. It physically interacts with its parent molecular cloud complex \citep[e.g.][]{Yoshiike_13,Reach_05,Wootten_77}.

In this paper, we focus on the need for spectral imaging of IC443 and W44 with good angular resolution and sensitivity at high-radio frequencies. This could help to better constrain the multi-wavelength scenario described above since local changes in the radio spectrum trace the actual energy distribution of the different electron populations responsible for both radio and part of gamma-ray emissions. This is of utmost importance for extended SNRs partially resolved by the angular resolution of gamma-ray instruments.
We provide radio mapping of IC443 and W44 at L (1.55 GHz) and C-band (7 GHz) obtained
by the recently commissioned Sardinia Radio Telescope (SRT\footnote{\textit{www.srt.inaf.it}}) during
Astronomical Validation (AV\footnote{\textit{http://www.srt.inaf.it/astronomers/astrophysical-validation-team/}}) and Early Science Program (ESP\footnote{\textit{http://www.srt.inaf.it/astronomers/early-science-program-FEB-2016/}}) activities.
We present accurate flux density measurements and spatially resolved spectral-slope
measurements, obtained through on-the-fly mapping procedures and using state-of-the-art 
imaging techniques. 

This work represents a first scientific milestone for SRT, testifying its capabilities and performances in single-dish imaging, and exploiting different receivers and backends.

\section{SRT observations}

SRT is a 64-m diameter radio telescope with
Gregorian configuration located on the Sardinia island (Italy), and is
designed to observe in the $0.3-116$ GHz frequency range. At present,
three receivers are available for observers: a 7-beam K-band receiver
($18-26.5$ GHz), a mono-feed C-band receiver ($5.7-7.7$ GHz), and a
coaxial dual-feed L-P band ($1.3-1.8$ GHz; $305-410$ MHz) receiver
\citep{Valente_16,Valente_10}. 
SRT offers advanced technology with the implementation of an active surface composed of 1008 panels and 1116 electromechanical actuators on the primary mirror. This allows us to compensate the gravitational deformations and to re-shape the primary mirror from a shaped configuration to a parabolic profile, depending on the position of the receiver at different antenna foci \citep{Bolli_15,Buttu_12,Orfei_04}. 

A first series of observations of  IC443 and W44 was performed during the AV of SRT, from May 27 to December 10 2014 ($\sim$12 hrs of effective time on targets). This phase was devoted to testing the performances of the telescope and the acquisition systems (Prandoni et al.,\,submitted). The targets were observed at 7.24 GHz 
with a bandwidth of 680 MHz.
%
Data were recorded with the Total-Power (TP) backend,
an analog device provided with
14 input chains, each of which brings signal to a
square-law-detector followed by a voltage-to-frequency converter to digitize the detected signals. Each of
these chains generates an output
that is proportional to the
amplitude of the incoming signal. The TP contains a corresponding number
of counters that count the impulses and integrate them for a programmable
integration time.
Subsequently, a program\footnote{ESP S0009, \textit{"Constraining cosmic ray production in SNRs with SRT"}, PI: A. Pellizzoni} dedicated to the observations of SNRs was approved during the Early Science phase of SRT, for a total exposure time on targets of $\sim20$ hrs in C and L-bands. 
The observations were performed in "shared-risk mode" at 7.0 GHz 
(bandwidth=1,200 MHz), and 1.55 GHz (bandwidth=460 MHz) between February 14 and March 24 2016, operating
simultaneously the TP and SARDARA (SArdinia Roach2-based Digital
Architecture for Radio Astronomy; Melis et al. in preparation) backends in
piggy-back mode. SARDARA is a new generation and flexible ROACH2-based
backend.
A total bandwidth of 1,500 MHz was available during ESP SARDARA
operations, split into $1,024\times2$ channels (left/right circular polarization) to
carry out most of our observations (part of L-band observations were performed with 16,384 channels).



The active surface was set in the shaped profile when observing in C-band in order to offer a better illumination of the Gregorian
focus, and in parabolic profile for L-band to increase the efficiency of the receiver installed at the primary focus.





We carried out the mapping of IC443 and W44 through On-the-Fly scans
(OTF). This technique implies that the data acquisition is continuously ongoing while the antenna performs constant-speed scans across the sky (typically a few degrees/min), alternatively producing maps along the Right Ascension (RA) and Declination (Dec) directions. In particular, the sampling time was set to 40\,ms during AV activities and 20\,ms during the ESP. A typical OTF off-source scan appears, in a signal-intensity-vs-time plot, as a linear slope, which represents the "baseline" (i.e. background emission and system-related signal).
The aforementioned parameters typically imply the acquisition of $>10-20$ samples/beam for each scan passage (largely oversampling the beam w.r.t. Nyquist sampling) allowing accurate evaluation of flux errors (see Section 3). In addition, beam oversampling allows us to efficiently reject outlier measurements ascribed to radio frequency interference (RFI).
The length of the scans is chosen according to the size of the source (typically of the order of $0.5-1^\circ$ for SNRs). In order to properly reconstruct the morphology of the observed source and its associated flux, the scan-dependent baseline (background flux) must be correctly subtracted. Ideally, each scan should be free from significant source contribution (and RFI contamination) for $40-60\%$ of its length/duration, in order to properly identify and subtract the baseline component. This requirement is not trivially satisfied for targets located in crowded regions of the Galactic Plane, as for the case of W44. 

The scan length corresponding to the observations of IC443 was set
to $1.5^\circ$ in both RA and Dec, accounting for the
size of the target of about $45\arcmin$ and baseline subtraction
requirements. Since W44 is a slightly less extended source ($\sim\,30\arcmin$), the scan length was set to $1.2^\circ$ in RA and
$1^\circ$ in Dec, accounting also for nearby sources contamination on
the Galactic Plane. Each scan duration was scheduled to 22.5 sec
for the first source, and 18 sec (RA) and 15 sec (Dec) for the second
one, which implies an OTF speed of $4\arcmin$/sec. Two consecutive
scans were separated by an offset of $0.01^\circ$, which implies
that 4.5 passages were carried out per beam on average, and about 17
samples per beam per scan (assuming a beam size of $2.7\arcmin$ in
C-band) were taken. The duration of additional dead time/slew time
varied between the observations, lasting for about $10-16$ sec at the
end of each scan (typically up to $30-40\%$ of overall
duration). The total duration of an observation (defined as a complete
map along both RA and Dec directions, including slew and dead time) at 7
GHz was about 3 hrs for IC443 and $\sim 2$ hrs for W44. The total
duration of a map at 1.55 GHz was much faster, $\sim 40$ min for IC443
and $\sim 30$ min in the case of W44.
Stable weather conditions (possibly a clear sky) have proved to be preferable for the production of high-quality maps, even at frequencies as low as the C-band one, as the baseline is visibly perturbed by the presence of an inhomogeneous cloud cover.
A summary of target  observations is reported in Table~\ref{tab:technical-obs}. 

A set of flux density calibration observations was performed through OTF cross-scans on standard point-like calibrators at the considered frequencies (3C286, 3C295, 3C123, 3C48, 3C147 and NGC7027) before and after each target map.




\begin{table}
	\centering
	\caption{Summary of the observations of IC443 and W44 performed with SRT during the AV and ESP. Freq, BW and HPBW indicate the central frequency, bandwidth and the Half Power Beam Width, respectively.}
	\label{tab:technical-obs}
	\begin{tabular}{lccccc} 
		\hline
			 	& Target	&	 Freq.		& BW			& HPBW	& Map size		 \\
			 	&  		& 	(GHz)		& (MHz)		& (arcmin)	&  ($^{\circ}$,$^{\circ}$)	\\	
		\hline
		AV &  IC443	& 7.24  		& 680	 	& $2.66 \pm 0.02 $	& $1.5\times1.5$	\\	
			     &  W44		& 7.24	  		& 680	& $2.66 \pm 0.02 $	& $1.2\times1.0$	\\	
		\hline
	ESP & IC443		&   7.00		& 1200	 	& $2.71 \pm 0.02 $		&  $1.5\times1.5$	\\
			   & 			&   1.55		&  460 	& $11.1 \pm 0.1$	& $2.0\times2.0$	\\
			   & W44		&   7.00		& 1200 	& $2.71 \pm 0.02 $	& $1.2\times1.0$	\\
			   & 			&   1.55		&  460 	& $11.1 \pm 0.1$	& $1.6\times1.4$ 	\\
		\hline
	\end{tabular}
\end{table}

\section{Imaging data analysis and calibration}

Data analysis was performed using the SRT Single-Dish Imager (SDI),
which is a tool designed to perform continuum and spectro-polarimetric imaging, optimized for OTF scan mapping, and suitable for all SRT receivers/backends \citep[see details in][]{Egron_16}.
SDI generates SAOImage DS9\footnote{http://ds9.si.edu} output FITS images suited to further analysis by standard astronomy tools. 
The core of our prodecure is to fully exploit the availability of a significant number of measurements per beam (and then per pixel, typically chosen to be about 1/4 of the HPBW), in order to have a straightforward evaluation of statistical errors (through standard deviation of the measurements in each pixel), efficient RFI outliers removal and accurate background baseline subtraction.

Our data analysis pipeline involves the following major steps/procedures:

\begin{table*}
	\centering
	\caption{
Flux densities of SNR IC443 for individual maps and related image parameters obtained with SRT during the AV and ESP.
 ESP observations at 7.0 GHz were simultaneously performed by TP and SARDARA backends in piggy-back mode.
The effective exposure time for each map (without deadtime) 
is also reported. "Total" refers to the total averaged map
parameters.}
	\label{tab:IC433-flux-rms}
	\begin{tabular}{lcccccccc} 
		\hline
\textbf{IC443}	 &	 MJD		& N. maps	& Eff. time	& 	Freq & 	Flux TP  &	rms TP &  Flux SARDARA  &  rms SARDARA  \\ 
			 &  			&(RA+Dec)			& (h,min)	& 	(GHz ) &		(Jy) &	(mJy/beam) & (Jy) &	(mJy/beam) 	\\ 
		\hline
	AV 	   & 56811		& 1				& 1h50 		&   	7.24	& $63.0 \pm 3.1$   &     22	 & -	& -	\\ 
			   & 56947		& 1.5				& 2h46 		&   		& $64.2 \pm 3.2$   &     11	 & -	& - \\ 
			   & 57001		& 1				& 1h50 		&   		& $73.4 \pm 3.7$  &     25	 & - & -	 \\ 
			   &\textbf{Total}& \textbf{3.5}		& \textbf{6h26}	 &   		&\textbf{63.6$^{\star}$ $\pm$ 3.1}    &   \textbf{6}  & - & -  \\ 
		\hline
		\hline
	ESP	  &\textbf{57440} & \textbf{3}		& \textbf{5h32}	 &   	7.00	& \textbf{66.9 $\pm$ 3.3} & \textbf{25} &\textbf{69.0 $\pm$ 3.5}    &  \textbf{20}  \\ 
		\hline
			   & 57432  	& 7.5			& 4h01 		&	1.55	 & -	& -	&$130.9 \pm 3.9$  &	88 	 \\ 
			   & 57469  	& 6				& 3h12 		&		 & -	& -	& $135.0 \pm 4.1$  &	58 	 	 \\
			   &\textbf{Total}& \textbf{13.5}		& \textbf{7h13}	 &   		& -   &  -    & \textbf{133.7 $\pm$ 4.0}    &   \textbf{76}  \\	
		\hline
	\end{tabular}\\
  	    \small
   	   $^{\star}$ The corresponding flux at 7.0 GHz is 64.4 Jy.
\end{table*}

1) {\it RFI rejection}. We provided an automatic and interactive RFI flagging procedure. Automated RFI detection and rejection is twofold. A "spectral RFI flagging" is based on automated search for outliers in each scan-sample's
spectrum (SARDARA data analysis only); for L-band, about 30\% of the channels are dynamically flagged and rejected
for suspected RFI contamination; for C-band, that percentage drops to $5-10\%$.
Spectroscopic SARDARA data are then collapsed into a single continuum channel after RFI spectral filtering.
A further "spatial RFI flagging" procedure (suitable for both Total Power and SARDARA backends) consists in splitting
the map into sub-regions, which correspond to adjacent solid angles in the sky. These areas have to be inferior to
the beam size (typically $1/4-1/5$ of HPBW) in order to avoid discarding actual fluctuations from the source, but
large enough so that they include a significant (typically >10) number of measurements. 
The "outlying" samples presenting a count level above a standard deviation-based threshold (typically $5\sigma$ level above average) are then flagged as RFI.

2) {\it Baseline subtraction}. 
Automated baseline subtraction is performed scan-by-scan.
A baseline "fitness" parameter (BF) is defined as the number/percentage of scan samples that are within a given rms (i.e. $1\sigma$ level of the baseline fluctuation) of a given baseline
linear fit. The higher the BF value, the better the accuracy of the baseline model. The BF parameter is maximized through a trial loop on the angular coefficient and normalisation of the linear fit. For example, linear fit trials performed on a scan on the empty sky, converge to a maximum $BF=68\%$ (i.e. $68\%$ of the samples are within $1\sigma$/rms from the best linear fit), as theoretically expected.  In principle, if at least $50\%$ of the scan samples are related to the baseline (i.e. if $<50\%$ of the samples are significantly contaminated by residual RFI and astrophysical sources), this procedure provides mathematically exact
results through BF maximisation: manual inspection/trimming of the baseline is not expected to provide more accurate (and rigorous) results.
The requirement of $50\%$ "source/RFI-free" for the scan samples cannot be trivially fulfilled on scans performed along crowded regions as -e.g.- the Galactic Plane.
An interactive final data inspection is then performed scan-by-scan in order to identify and adjust anomalies in baseline
subtraction and RFI rejection (e.g. discarding corrupted scans, further manual flagging/unflagging of RFI).

3) {\it Calibration}. For each OTF cross-scan performed on calibrators, after automatic subtraction of the baseline, our pipeline
applies a Gaussian fit to the data in order to measure the calibrator scan peak counts and check for pointing errors.
For L-band, pointing errors are negligible, while for C-band they sporadically range up to $0.5\arcmin$ 
(i.e. exceeding the SRT pointing requirement of HPBW/10), thus we need calibrator counts correction up to $\sim$10\%.

The spectral flux density of the calibrators at the observed frequency 
was reconstructed/extrapolated from the values and the polynomial
expressions proposed by \citet{Perley_13} 
using the VLA data. We note that among our set of observed calibrators, 3C123, 3C286 and 3C295 show flux variations of less than $5\%$ per century between 1 and 50 GHz. 
Differences among calibrators and SNR spectra on a relatively large bandwidth could in principle affect calibration results.
However, we verified that bandpass corrections applied to SARDARA and TP data affect calibration results below $\sim$0.3\%.
The conversion factors Jy/counts for both left and right circular polarization channels (obtained from the ratio between calibrator
expected flux density and observed counts) was established for each calibrator observation at different elevations by averaging the
values associated with consecutive cross-scans.
Calibration factors are roughly independent from the elevation since both L-band and C-band SRT gain curves are approximately flat due to 
optimised settings of the antenna's active surface (parabolic and shaped mode, respectively).
We discarded calibrators and target observations obtained at elevations below 15$^{\circ}$ for C-band observations since pointing errors and beam shape instability effects are appreciable at low 
elevations.
Calibration factors are grouped and averaged before being applied to
scan samples according  to the following criteria: 
(1) calibration and target data must be observed with the same backend attenuation parameters; (2) only calibration
observations performed within 12 hrs (or less in case of changing weather) from each scan epoch are considered
in order to  guarantee the same conditions of target/calibrators observations and the highest stability of the conversion factor.
Following the above procedure, the gain stability and related errors (i.e. standard deviation on
the calibration factors) are $\sim 3\%$  for L-band and up to $\sim 5\%$ for C-band (depending 
on actual weather conditions for C-band).

In the considered elevation range $\sim 20-80^{\circ}$ the antenna beam has proved to be very stable (HPBW $= 2.71\arcmin \pm 0.02\arcmin$ at 7.0 GHz and HPBW $= 11.1\arcmin \pm 0.1\arcmin$ at 1.55 GHz).
A Gaussian shape provides a very good fit to OTF scans on calibrators,
thus we assumed a beam solid angle of $\pi$(1.2$\times$HPBW/2)$^2$ sr for image calibration. The HPBW was weighted accordingly to target spectra in the observed band.


\begin{table*}
	\centering
	\caption{Same as in Table~\ref{tab:IC433-flux-rms} for SNR
          W44.}
	\label{tab:W44-flux-rms}
	\begin{tabular}{lcccccccc} 
		\hline
\textbf{W44}	 &	 MJD		& N. maps	& Eff. time	&	Freq &	Flux TP & rms TP  &	Flux SARDARA &	rms SARDARA \\
			 &  			&(RA+Dec)			& (h,min)			&  (GHz )  &	(Jy) &	   	(mJy/beam) & (Jy) &	(mJy/beam)	\\	
		\hline
	AV	   & 56847  	& 1				&  0h59 		&	7.24	& $92.2 \pm 4.6 $&	24 	& -	& - \\
			   & 56911		& 3				&  2h56		&   		& $90.7 \pm 4.5 $ &     16 	& -	& - \\	
			   & 56916		& 1				&  0h59		&   		& $92.7 \pm 4.6 $ &     16 	& -	& - \\	
			   &\textbf{Total}& \textbf{5}		& \textbf{4h54}	 &   		&\textbf{91.4$^{\star}$ $\pm$ 4.6}   & \textbf{13} & -& -\\
		\hline
		\hline

	ESP	   & 57441  	& 2				& 1h57		&	7.00	 & $92.0 \pm 4.6$	& 12	&	$93.9 \pm 4.7$   &	15 	 \\
			   & 57442  	& 5				& 4h53 		&		 & $93.1 \pm 4.7$	& 15	&	$96.9 \pm 4.8$  &	10 	\\	
			   &\textbf{Total}& \textbf{7}		& \textbf{6h50}	 &   		&  \textbf{92.9 $\pm$  4.6}	& \textbf{11}	&	\textbf{95.7 $\pm$ 4.8}    &  \textbf{7}    \\	
		\hline
 			   & 57432  	& 5				& 1h30		&	1.55	 & -	& -	&	$214.8 \pm 6.4$  &	81   \\
			   & 57433  	& 4				& 1h11 		&		 & -	& -	&	$210.3 \pm 6.3$  &	92 	 \\
			   & 57470  	& 5				& 1h30 		&		 & -	& -	&	 $215.3 \pm 6.5$	 &	85  \\
			   &\textbf{Total}& \textbf{14}		& \textbf{4h11}	 &   		& -	& -	&	\textbf{214.4 $\pm$ 6.4}   &   \textbf{81}   \\	
		\hline
	\end{tabular}\\
  	    \small
   	   $^{\star}$ The corresponding flux at 7.0 GHz is 92.5 Jy.
\end{table*}

4) {\it Image production}. Calibrated data were binned through ARC tangent projection using pixel sizes $\sim$1/4 of the HPBW, which corresponds to the effective resolution of the images. Note that bright and nearby point-like image features (e.g. having $>0.1$ Jy flux and a beam-size separation) associated with a Gaussian Point Spread Function (i.e. SRT beam shape) are not distinguishable when taking an image pixe size equals to the  HPBW, while these are resolved when adopting a pixel size equals to the effective resolution ($\sim 1/4$ HPBW). This arises from Gaussian beam oversampling in our mapping procedures.

DS9 FITS images were produced in units of Jy/beam and Jy/sr by incorporating the WCS reference system. The images are suitable for scientific analysis
with SAOImage (image rms, dynamic range, brightness profiles etc.). The astrometric accuracy was cross-checked to be at least an order of magnitude
better than the beam size by mapping the calibrators 3C295 and 3C286.
Due to the applied image oversampling ($>$100 measurements/beam on cumulative maps), statistical errors on flux density measurements
are straightforwardly obtained by calculating the standard deviation for each pixel. Integrated statistical flux errors (typically $<$0.5\%) arise from propagation of pixel
errors over the whole source extent and are well below systematic errors.
Systematic errors on absolute flux density measurements arising from
residual errors (after the corrections described above) on pointing accuracy, bandpass,
baseline subtraction, beam model uncertainties, receiver/backend linearity and stability are estimated to be cumulatively below $\sim$3\%.
%

We analysed each RA/Dec scan separately in the different maps in order to check for time-dependent residual RFI, baseline subtraction problems and other possible instrumental anomalies. 
Scans affected by major calibration problems (e.g. patchy opacity in C-band due to dense clouds) and/or low signal/noise were discarded before merging
the whole data sample in order to optimise final image accuracy and minimise the image rms. 
C-band calibrated images were produced by TP and SARDARA backends separately, while only SARDARA calibrated maps were processed in L-band, since low-frequency TP maps are affected by unrecoverably strong RFI contamination that can be discarded only through spectral filtering.

\section{Results}

We measured the integrated flux densities of our targets by defining a centroid of the diffuse emissions and considering a radius of $0.5^\circ$ for IC443 (centroid coord: $\alpha$=06h16m58s, $\delta$=$22^{\circ}31.6\arcmin$), and of $0.32^\circ$ in the case of W44 (centroid coord: $\alpha$=18h56m05s, $\delta$=$01^{\circ}21.6\arcmin$). 
We note that the particular choice of the extraction radius does not significantly affect flux measurements when including the whole SNR and excluding external sources, since baseline-subtracted off-source pixels have zero mean flux.

\begin{figure*}
     \includegraphics[width=1.0\textwidth]{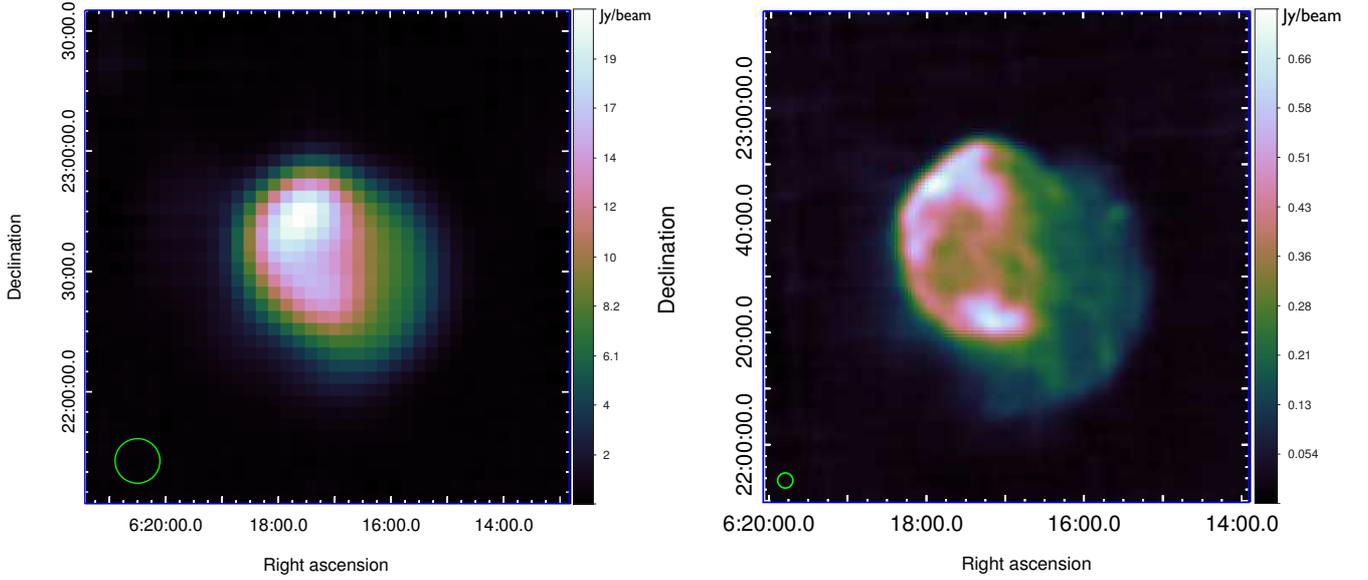}
    \caption{Left: SRT continuum map of SNR IC443 obtained at 1.55 GHz with the SARDARA backend
      during the ESP. Right: Continuum map of the 7.2 GHz
      observations performed with the TP backend during the AV tests. The green circles
      on the bottom  left indicate the beam size at the observed frequencies. Pixel sizes are $3.0\arcmin$ and $0.6\arcmin$ for 1.55 GHz and 7.2 GHz maps, respectively,
which corresponds to 1/4 HPBW.}
    \label{fig:map-IC433-L-C}
\end{figure*}

\begin{figure*}
	\includegraphics[width=1.0\textwidth]{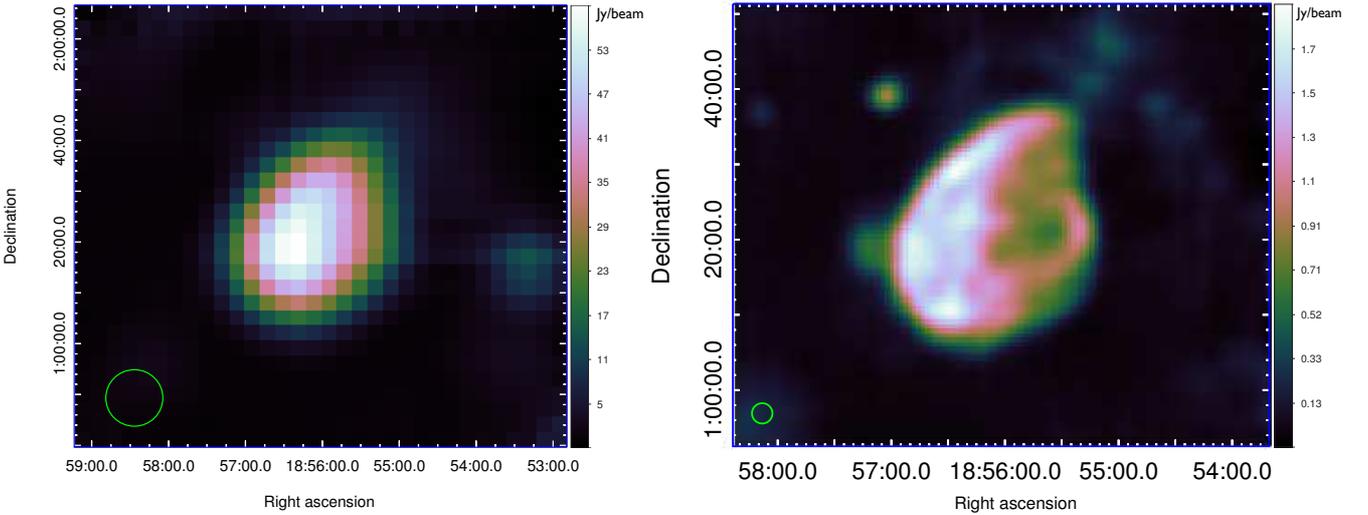}
    \caption{SRT continuum maps of SNR W44 obtained with the SARDARA backend at 1.55 GHz (left) and 7.0 GHz (right) during the ESP. The green circles on the bottom left indicate the beam size at the observed frequencies. Pixel sizes are as in Fig.\,1.}
    \label{fig:map-W44-L-C}
\end{figure*}

Individual and total averaged map parameters of IC443 and W44, resulting in merging the selected data sets,
are reported in Tables~\ref{tab:IC433-flux-rms} and \ref{tab:W44-flux-rms}, respectively, together with integrated flux densities.
The values obtained at 7.0 GHz using simultaneously the TP and SARDARA (operating in
piggy-back mode) during the ESP are fully consistent
and also compatible with AV measurements performed at 7.24 GHz more than one year earlier  within $1\sigma$. 
Averaging these results, we obtain a 7.0 GHz continuum flux of $66.8\,\pm\,2.9$ Jy for IC443 and $93.7\,\pm\,4.0$ Jy for W44.
The integrated fluxes measured with SARDARA at 1.55 GHz are 
$133.7\,\pm\,4.0$ Jy for IC443 and $214.4\,\pm\,6.4$ Jy in the case of W44.
We note that the rms associated with the maps is slightly better in the case of
SARDARA with respect to the TP when observing with identical weather conditions, as expected from a new-generation spectroscopic backend that allow us a better RFI rejection.
The total averaged maps of IC443 and W44 obtained at  1.55 GHz and 7 GHz are presented in Fig.~\ref{fig:map-IC433-L-C} and \ref{fig:map-W44-L-C}, respectively.

The mean surface brightness associated with the
brightest rims of IC443 is $\sim 0.6$ Jy/beam at 7 GHz, which is about five times larger than that corresponding to the western halo. In the case of W44, the brightest
rims have a mean brightness of $\sim 1.8$ Jy/beam.

The map of W44 also includes a source
at ($\alpha, \delta$)=(18h57m04, 1$^{\circ}38\arcmin45\arcsec$), which is also visible in VLA images at low frequencies and other wavelengths (IR with Spitzer). 
It could be associated with 
a classified HII region G035.040-00.510, as listed for example in the WISE Catalog of Galactic HII regions\footnote{astro.phys.wvu.edu/wise/}.
The corresponding flux density is $0.94\,\pm\,0.05$ Jy at 7.2 GHz. 
The associated spectral index (obtained through C-band and L-band point-like flux densities) is very hard: $\alpha=-1.1\,\pm0.1$ (with $S_{\nu} \propto \nu^{-\alpha}$).

\begin{figure*}
	\includegraphics[width=1.0\textwidth]{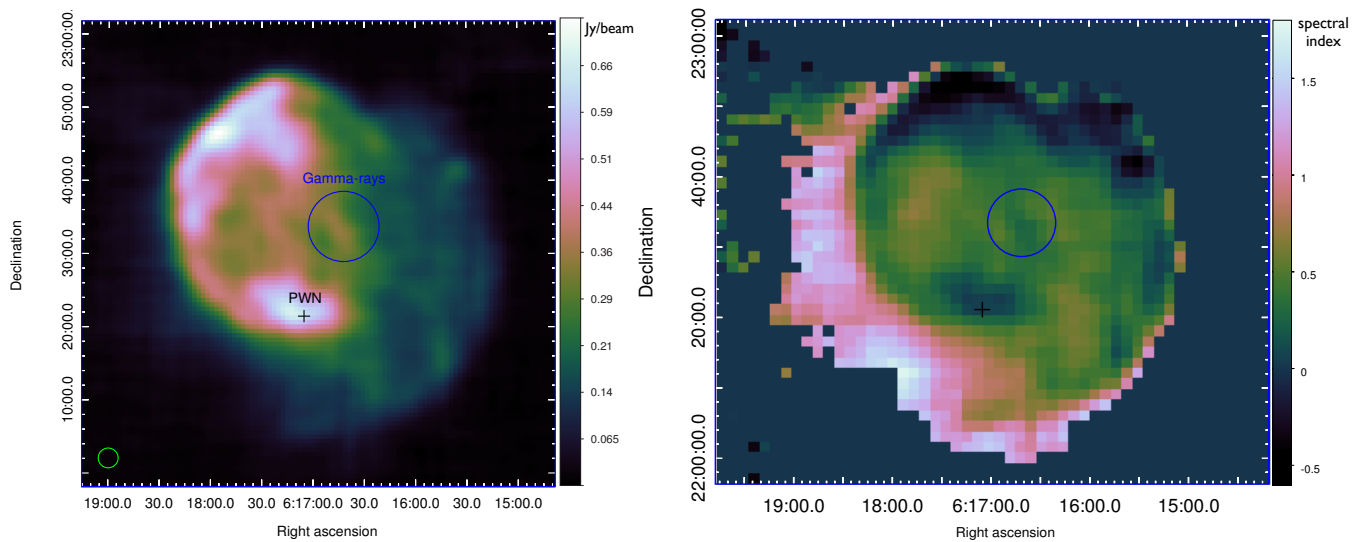}
    \caption{Continuum map of SNR IC443 at 7 GHz (left) and spectral index map obtained by using 1.55 and 7 GHz data (right).
The black plus simbol indicates the position of the PWN, whereas the blue
circle indicates the bulk of the gamma-ray emission seen with VERITAS \citep{Humensky_15}.
}
    \label{fig:IC443-spectral-index-map}
\end{figure*}

\begin{figure*}
	\includegraphics[width=1.0\textwidth]{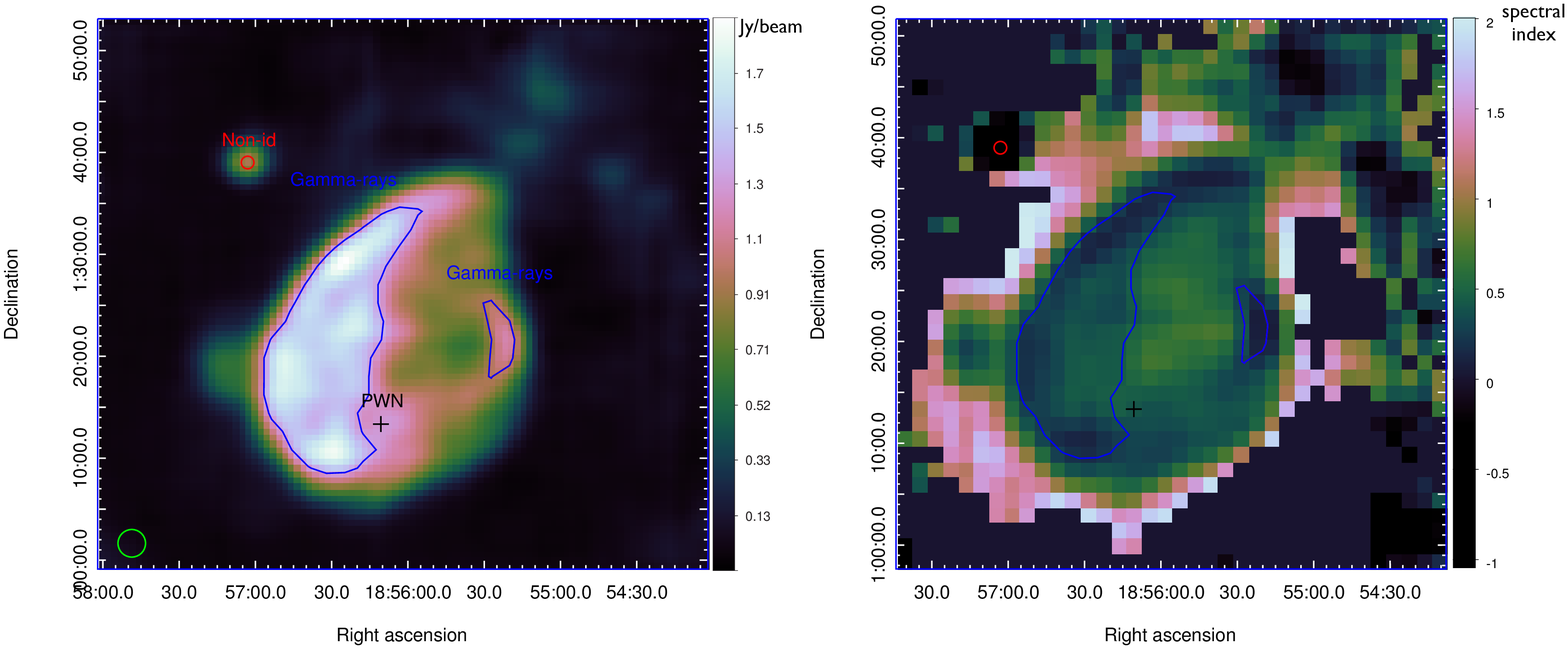}
    \caption{Continuum map of SNR W44 at 7 GHz (left) and spectral index map obtained by using 1.55 and 7 GHz data (right).
The black cross and the red circle indicate the position of the PWN
and the non-identified source, respectively. The blue areas
correspond to the gamma-ray emissions detected with AGILE and Fermi-LAT \citep{Giuliani_11B,Abdo_10}.}
    \label{fig:W44-spectral-index-map}
\end{figure*}

Comprehensive spatially-resolved spectra were determined by combining C-band and L-band maps produced with the same parameters and topology (same projection centre and pixel size).
In particular, spectral index maps were obtained by convolving data at low resolution (accordingly to L-band resolution).
The corresponding images are reported in Fig.~\ref{fig:IC443-spectral-index-map} and \ref{fig:W44-spectral-index-map}
for IC443 and W44, respectively, together with the surface brightness maps of the sources in C-band.
Since on-the-fly single-dish imaging techniques allow us precise measurements of the
density flux for each direction in the sky (i.e. for each pixel map),
the spectral index estimate is directly obtained fitting the two L-band and
C-band measurements for each pixel. Single-dish flux measurements are
straightforward with respect to the more complex interferometric data processing,
and related errors are typically lower (conservatively, about $3\%$ in L-band). On the
other hand, the available angular resolution for low-frequency single-dish imaging is much worse
than that from interferometric images.
Our first low-resolution result for single-dish spectral mapping of SNRs gives us an estimate of region-dependent spectral indices and a direct correlation with the corresponding brightness of the largest sub-regions targets (small structures are missing in our spectral index maps).
More accurate (i.e. higher angular resolution) images could be provided for example through combination of C-band and K-band (22 GHz) images, the latter having about $40\arcsec$ HPBW.

For both sources we observed a significant spread in the distribution of the spatially-resolved spectra ranging from a flat (for W44) or slightly inverted (up to -0.5 for IC443) spectral index to a relatively steep spectrum ($\alpha$$>$0.7) compared to a mean value of $\sim$0.5.
Relative errors among adjacent pixels are $<$1\%.
Our data highlight a correlation between the brightest flux density regions and flat spectral indices for IC443 and W44.
These regions are mainly located on the edge and on the brightest filaments of the SNRs. 
They match with the bulk of the gamma-ray emission in the case of W44. The approximate location of the gamma-ray emission detected with VERITAS and Fermi-LAT for IC443 \citep{Humensky_15}, and with AGILE \citep{Giuliani_11B} and Fermi-LAT in the case of W44 \citep{Abdo_10}, was indicated in blue in the Fig.~\ref{fig:IC443-spectral-index-map} and \ref{fig:W44-spectral-index-map}).

\section{Discussion}

We obtained accurate images of W44 and IC443 probing the capabilities of 
the Sardinia Radio Telescope by operating single-dish OTF scans.
Our results show reliable performances of the instrumentation over a two-year-long time span 
(gain stability $<$5\%)  and provide self-consistency checks
in measurements performed by simultaneous piggy-back observations with
two different backends.
On the other hand, this early SRT mapping of SNRs yielded challenging scientific results, providing the first spatially-resolved spectra in the $1.5-7$ GHz range.

Our results are first compared with both low-frequency VLA interferometric data and
existing $0.8-5$ GHz single-dish maps (\S 5.1), and related integrated fluxes described in the literature (\S 5.2).
In \S 5.3 we discuss the physical implications of the observed region-dependent spectral indices for IC443 and W44 and their correlation with the radio and gamma-ray intensity maps.








\subsection{SRT single-dish imaging performances}

Interferometry can "pass the baton" to single-dish techniques to image large structures ($\sim 1^{\circ}$ or more) at high frequencies, since synthesis imaging becomes difficult in this context.
In fact, the most prominent features highlighted
in the SRT maps at 7 GHz are coherent with the high-resolution, low-frequency interferometric images of IC443 and W44.
Thus, good mapping quality of large structures can be maintained
even at high-frequencies if deep and oversampled OTF single-dish mapping is provided.

It is worth noting that the SRT image at 7 GHz can better assess the actual SNR edges (and then also SNR fluxes) than other single-dish maps that are 
strongly affected by the contamination of radio emission from nearby sources in the crowded regions of the Galactic Plane, as in the case of W44.

\subsubsection{IC443}

\begin{figure*}
	\includegraphics[width=\textwidth]{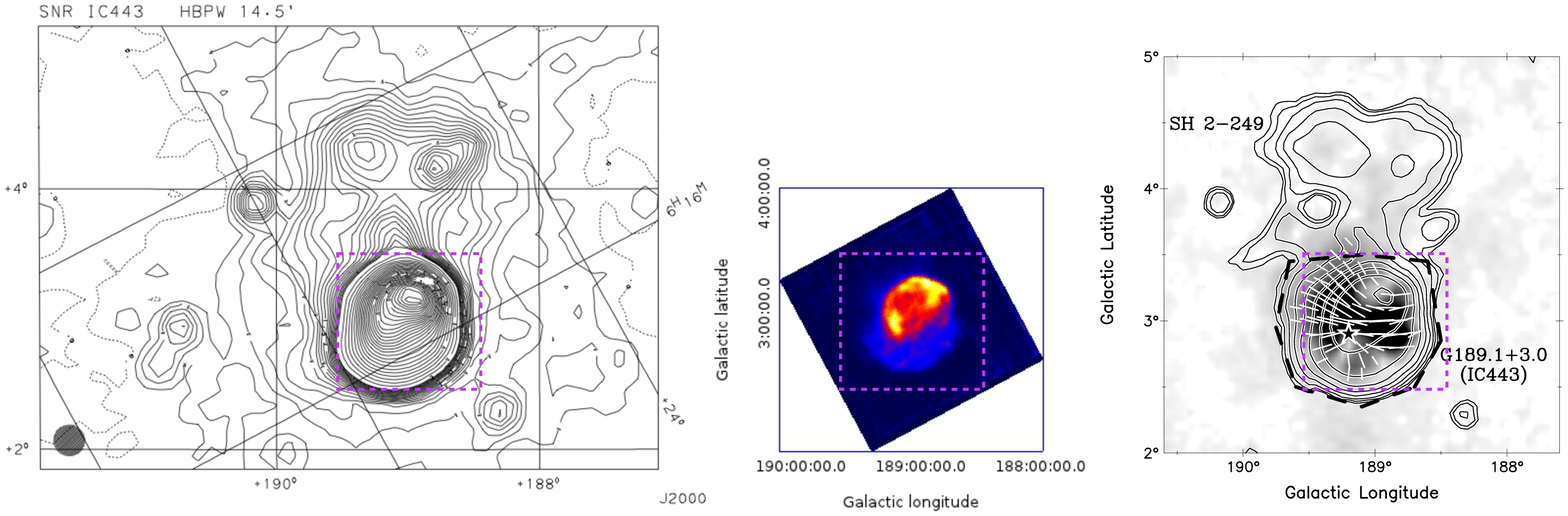}
    \caption{Comparison of the radio continuum emission of SNR IC443 as seen  with different single-dish telescopes. Left: Contour plot observed at 868 MHz with Effelsberg \citep{Reich_03}. Middle: SRT observation at 7.2 GHz in Galactic coordinates. 
Right: Radio continuum and polarization image of IC443 obtained with Urumqi at 5 GHz
\citep{Gao_11}. The dotted rectangles indicate the same sky region. 
}
    \label{fig:IC433-Eff-Urumqi-SRT}
\end{figure*}

Single-dish observations of IC443 were performed with Effelsberg at 868 MHz
in $1999-2000$ \citep{Reich_03}, and with Urumqi during a 5 GHz
polarization survey of the Galactic Plane between 2004 and 2009 \citep{Gao_11}. The resulting images are compared with the map obtained with SRT at 7 GHz in Fig.~\ref{fig:IC433-Eff-Urumqi-SRT}. The three observations highlight the presence of different, strong intensity regions in the SNR. The interaction of IC443 and the HII region (S249) located at the north (Galactic latitude) of the remnant is clearly visible on the maps obtained with Effelsberg and Urumqi, while SRT offers more details on the morphology of the supernova remnant, thanks to the better angular resolution of $2.7\arcmin$ with respect to Urumqi (HPBW=$9.5\arcmin$) and Effelsberg (HPBW=$14.5\arcmin$) at the observed frequencies.

\begin{figure*}
         \centering
	\includegraphics[width=0.9\textwidth]{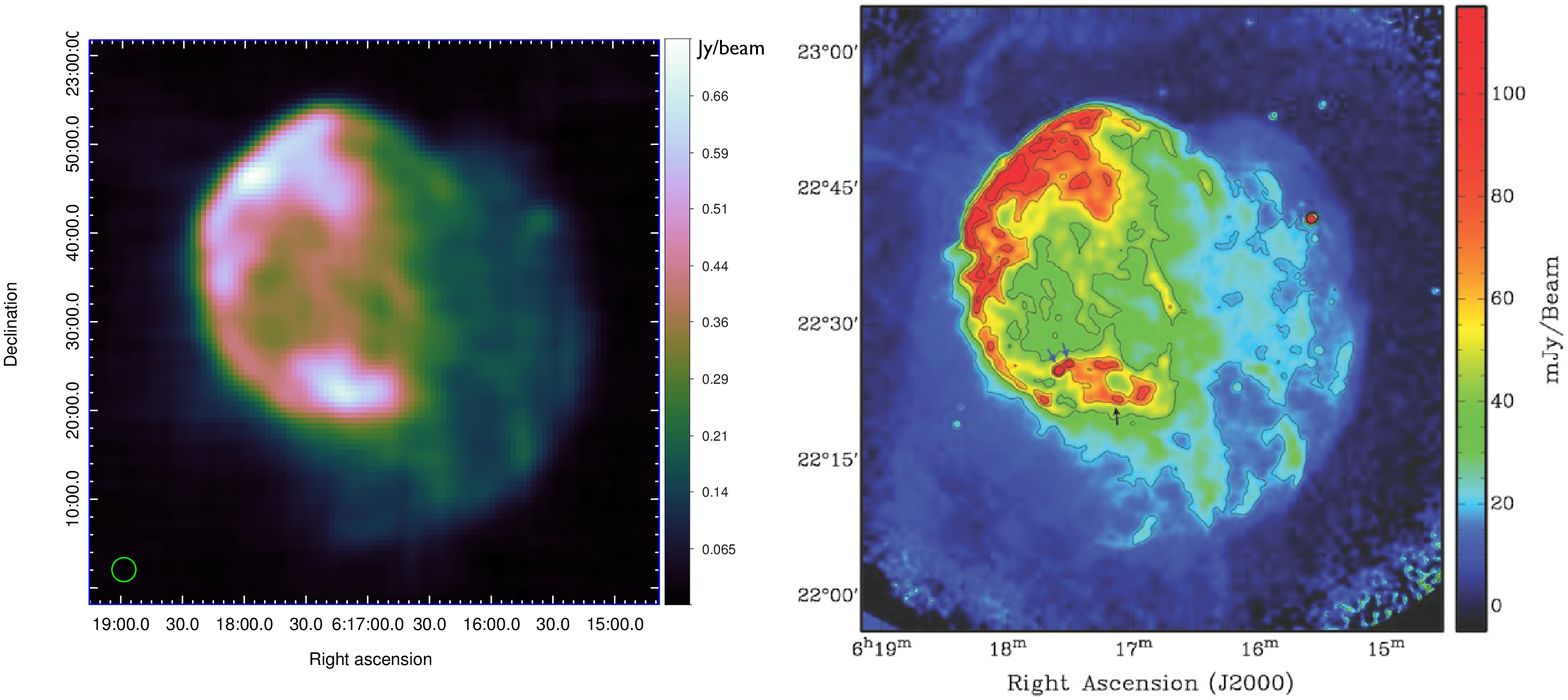}
    \caption{Comparison of the continuum maps of IC443 obtained with SRT at 7 GHz (left) with that obtained combining VLA and Arecibo at 1.4 GHz
(right) \citep{Lee_08}.}
    \label{fig:IC433-VLA-Arecibo}
\end{figure*}

We then compared the SRT map with high-resolution maps of IC443
obtained by coupling interferometric and large-area single-dish observations.
VLA and Arecibo low-frequency data at 1.4 GHz were combined together to achieve an extremely good sensitivity and angular resolution of about $40\arcsec$ \citep{Lee_08}. 
%
The main features of the morphology of IC443 revealed with SRT at 7 GHz are
consistent with those in the above observations,
as testified by Fig.~\ref{fig:IC433-VLA-Arecibo}, and with the image obtained at 330 MHz (not shown in this paper) with the VLA \citep{Castelletti_11}.

IC443 consists in two nearly concentric shells, which present a complex structure with the presence of filaments (not resolved, but detected in the SRT image), and a clear difference in the radio continuum intensity.
The eastern shell is open on the western side toward a weaker second shell (halo), which is possibly related to a breakout portion of the supernova remnant into a rarefied medium \citep{Lee_08}. 
The eastern shell shows two very bright emission regions, which are apparently connected with a ridge.
The bulk of the emission comes from the northeastern part of this shell, corresponding to signatures of atomic/ionic shock \citep{Duin_75}.
The bright emission at the southwestern part of the ridge has a more complex origin.
Various signs of molecular shock of H2 were highlighted by \citet{Burton_88}. 
The pulsar and its PWN 
are located in this region. They very likely correspond to the
remnant of the explosion which later formed IC443. We note that no
pulsations were detected from the neutron star, however, all
evidence points to the nature of a rotation-powered pulsar \citep[][and
references therein]{Swartz_15}. 
Two bright extragalactic point-like sources unrelated to
the remnant \citep{Braun_86} are also present near
($\alpha,\delta$)=(06h17m30s, $22^{\circ}25\arcmin$),
although these are not resolved in the SRT map.


\subsubsection{W44}

We compared the SRT map of W44 with radio continuum maps obtained by Effelsberg at 4.9 GHz and Urumqi at 4.8 GHz,
and with VLA low-frequency observations.

\begin{figure*}
	\includegraphics[width=\textwidth]{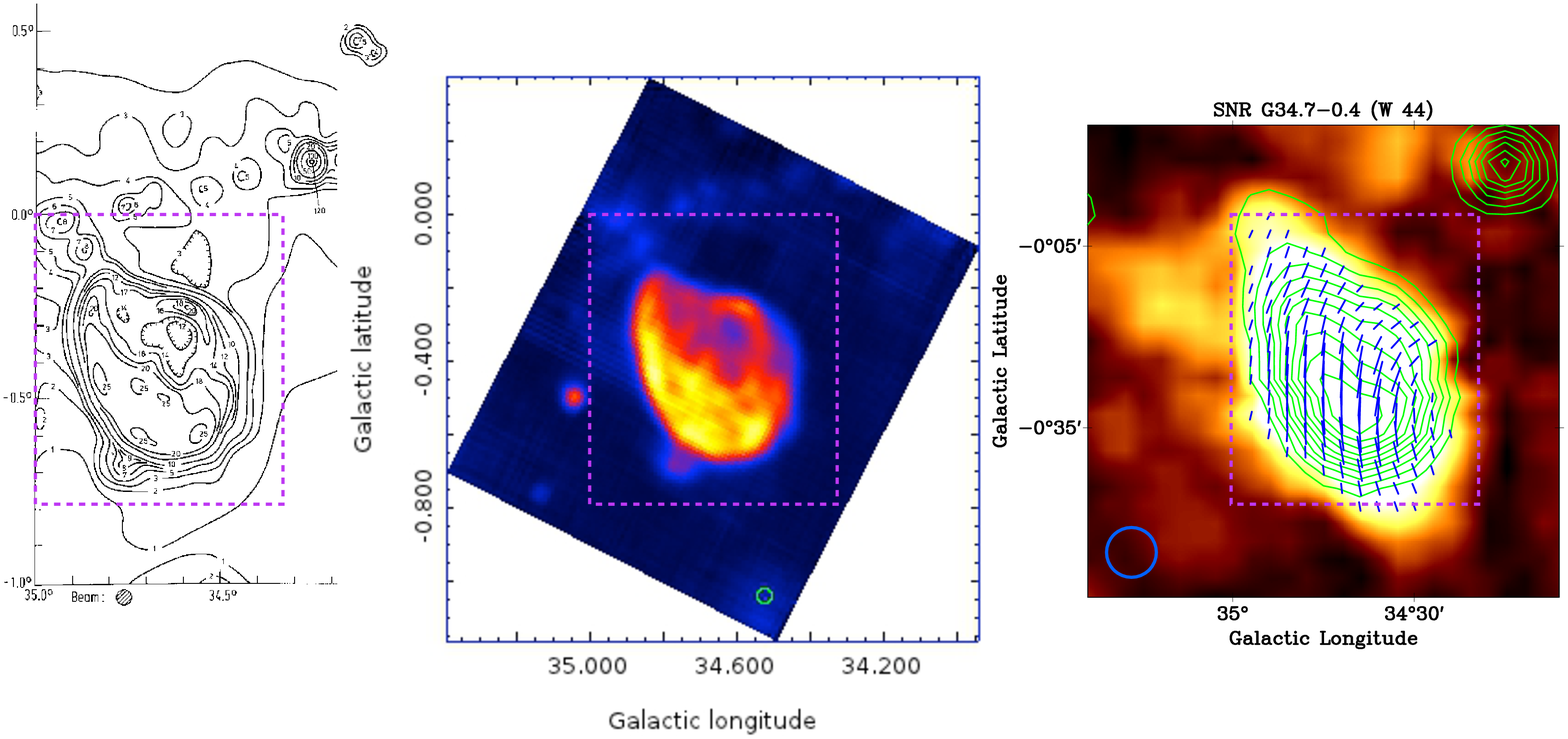}
    \caption{Left: contour map obtained by
      Effelsberg at 4.9 GHz \citep{Altenhoff_79}, middle: SRT observation at 7 GHz, right: intensity contours (indicated in green) obtained with Urumqi at 4.8 GHz \citep{Sun_11}. The dotted  rectangles indicate the same sky region. The blue and green circles in the bottom corner of the maps show the beam size.}
    \label{fig:W44-Eff-Urumqi}
\end{figure*}

A survey of the Galactic Plane was produced by the Effelsberg telescope at 4.875 GHz \citep{Altenhoff_79}. Scans were taken 
in galactic latitudes over $b= \pm2^{\circ}$, at a rate of $80\arcmin$/min, and spaced every $1\arcmin$ in galactic longitude. 
The HPBW was $2.6\arcmin$, which gives us a direct comparison with the SRT
resolution at 7.2 GHz ($2.7\arcmin$). The maps obtained with both radio telescopes
 are very similar, as shown in Fig.~\ref{fig:W44-Eff-Urumqi}. SNR W44 presents strong intensity regions that are located mainly in 
the south (Galactic latitude) of the remnant. It is worth noting the emission from the Galactic Plane.
A more recent observation of W44 was performed during a polarization survey carried out with Urumqi at 4.8 GHz \citep{Sun_11}. Details of 
the supernova remnant, the Galactic Plane, and sources nearby W44 are clearly visible with SRT. Instead, W44 appears more
extended in the case of Urumqi, since the sources in the vicinity of W44 and part of the Galactic Plane are not resolved from the remnant (see Fig.~\ref{fig:W44-Eff-Urumqi}), as for example in the case of the unidentified source detected at $\sim$1 Jy by SRT in the north-east direction. 
This is related to the beam width associated with Urumqi at 4.8 GHz, which is about 3.5 times larger than that of SRT at 7 GHz 
($9.5\arcmin$ for Urumqi against $2.7\arcmin$ for SRT).

\begin{figure*}
         \centering
    \includegraphics[width=0.8\textwidth]{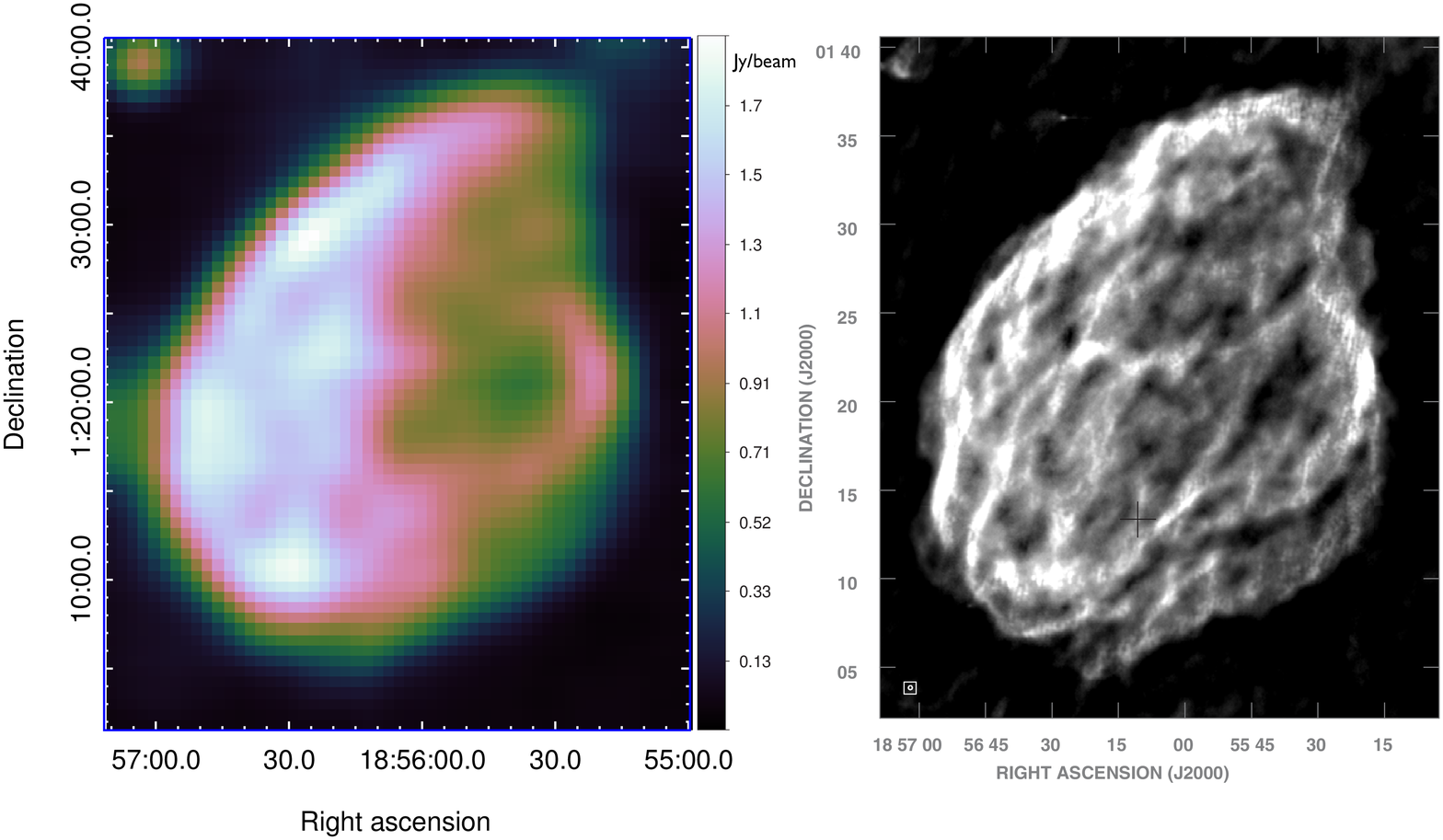}
    \caption{Comparison of the continuum maps of SNR W44 obtained with SRT at 7.0 GHz (left) with that obtained with the VLA at 324 MHz (right) \citep{Castelletti_07}. The beam size in the SRT map is represented by the point-like source in the top left corner.}
    \label{fig:W44-VLA}
\end{figure*}

We then compared the map of W44 produced by SRT with high-resolution
VLA images of the remnant at 1,465 MHz \citep{Jones_93} and 324 MHz \citep{Castelletti_07}, 
which were obtained using interferometric multiple-configurations (see
Fig.~\ref{fig:W44-VLA} for a comparison with the 324 MHz map). 
The radio emission of W44 is characterized by an asymmetric limb-brightened shell. 
%
The brightest filaments are also detected with SRT (although not resolved), and they most likely result from radiative shocks driven into clouds or sheets of dense gas \citep{Jones_93}.
The brightest emission occurs along the eastern boundary at $\sim (\alpha, \delta)$=
(18h56m50s,$01^{\circ}$17$\arcmin$).
It results from the interaction between W44 and dense molecular clouds observed in this region \citep{Seta_04,Reach_05}. 
\citet{Castelletti_07} using Spitzer observations at 24 $\mu$m and 8 $\mu$m identified a circular HII region centered at ($\alpha,\delta$)=(18h56m47.9s, $01^{\circ}$17$\arcmin$54$\arcsec$) 
and named G034.7-00.6 \citep{Paladini_03}, with the IRAS point source 18544+0112, which is a young stellar object located on its border.
This feature appears at about 0.7 Jy/beam in the SRT 7 GHz image at ($\alpha,\delta$)=(18h57m10s, $01^{\circ}$18$\arcmin$).
To the west, a short bright arc is visible at ($\alpha,\delta$)=(18h55m20s, $01^{\circ}$22$\arcmin$) \citep{Castelletti_07}. 
It corresponds to the SNR shock colliding with a molecular cloud located in this region, which is consistent with bright optical filaments \citep{Mavromatakis_03,Giacani_97} and IR observations.
The radio and X-ray nebula associated with the pulsar \citep{Petre_02,Frail_96}
has a slight offset w.r.t. the pulsar and is consistent with a motion of the pulsar away from the SNR center \citep{Jones_93}.

\subsection{IC443 and W44 continuum fluxes}

We compared the values of the integrated flux densities obtained
with SRT at 1.55 GHz and 7 GHz with those presented in the literature.
We note that no recent observations above 2 GHz were performed since
the late 1970's.

Based on the  values in Table 2 by \citet{Castelletti_11}, we conclude that our measurements for IC443 ($S_{\mathrm{1.55GHz}} = 133.7\,\pm\,4.0$ Jy and $S_{\mathrm{7GHz}} = 66.8\,\pm\,2.9$ Jy) are consistent with continuum fluxes reported at 1.4 GHz
($S_{\mathrm{1.4GHz}} = 130\,\pm\,13$ Jy originally obtained by \citealt{Green_86} then scaled accordingly to \citealt{Baars_77} flux density scales) and at
6.6 GHz  ($S_{\mathrm{6.6GHz}} = 70\,\pm\,15$ Jy
by \citealt{Dickel_71}), 
within $1\sigma$.
It is worth noting that typical continuum flux errors for IC443 in the literature are of the order of $\sim 10-15\%$, while we provided more accurate measurements. This is mostly due to our oversampled maps in which, for each pixel, tens of OTF baseline-subtracted scans are available, providing straightforward error measurements through standard deviation estimates.
From a weighted fit of the SRT data (Table 2), our spectral index estimate for IC443 in the interval $1.55-7.2$ GHz is $\alpha=0.46 \pm 0.03$.
This result is in perfect agreement with a spectral index estimate of $\alpha=0.47 \pm 0.06$ 
obtained through a weighted fit of all previous radio data available in the literature considering the frequency range $1.39-8.0$ GHz (see Table 2 in \citealt{Castelletti_11} for a direct comparison). 
Note that SRT measurements halve the spectral index error previously obtained for the high frequency range.
We focused the same analysis on the literature data at low frequencies, 
excluding the very low frequency data at 10 MHz which are possibly affected by a turnover due to thermal absorption as suggested by \citet{Castelletti_11}.
We obtained that the IC443 low frequency spectrum ($0.02-1.0$ GHz) flattens to  $\alpha = 0.33 \pm 0.01$.
Thus, a slight steepening of the IC443 spectrum ($\mathrm{\Delta} \alpha\,\sim\,0.1$) around $\sim$1 GHz could be speculated considering previous literature data (without including SRT measurements), though not statistically  significant: $\sim2\sigma$ level. This conjecture is now supported at $> 4\sigma$ confidence level using SRT observations alone, thanks to the smaller error on the high-frequency spectral index.

For W44, the continuum flux results presented in the literature, corrected by \citet{Castelletti_07} to \citet{Baars_77} flux density scales when possible, provide a wide scatter.
In particular, measurements at 1.4 GHz by \citet{Giacani_97}
($S_{\mathrm{1.4GHz}}=210\,\pm\,20$ Jy) and \citet{Castelletti_07}
($S_{\mathrm{1.4GHz}}=300\,\pm\,7$ Jy)
are inconsistent. Our result at 1.55 GHz ($S_{\mathrm{1.55GHz}}=214.4\,\pm\,6.4$ Jy) is
comparable within $1\sigma$ with the former and most other 
L-band flux measurements obtained in 
the '60s \citep{Beard_69,Pauliny-Toth_66,Scheuer_63,Leslie_60}.
Our measurement at 7.0 GHz ($S_{\mathrm{7GHz}}=93.7\,\pm\,4.0$ Jy) is in agreement with \citet{Hollinger_66}: $S_{\mathrm{8.3GHz}}=95\,\pm\,23$ Jy.

W44 displays a similar spectral behaviour as compared to IC443. 
In fact, the high-frequency spectral index estimated by SRT for W44 is $\alpha = 0.55 \pm 0.03$ ($1.55-7.2$ GHz).
%
Looking at low-frequency flux measurements reported in Table 2 of \citet{Castelletti_07}, the 
resulting spectral index is significantly softer. 
A weighted fit to the data in the $0.02-1.0$ GHz range  gives a spectral index $\alpha = 0.36 \pm 0.02$ associated with a very significant goodness-of-fit (reduced $\chi^{2}=1.06$).
Thus, SRT data suggest a possible spectral steepenening at high frequencies for IC443 and W44.
In Fig.\,9 we reported the spectra related to the data reported in the literature\footnote{Errors on \citet{Altenhoff_70} flux values are not correctly reported in the Table 2 by \citet{Castelletti_07}, as also noticed by \citet{Onic_15}. We included in Fig.\,9 the actual errors (10\%) at 1.4, 2.7 and 5 GHz.}
 and SRT measurements.
More observations are required in the $1-10$ GHz range in order to confirm these spectral turnovers and better constrain their parameters (e.g. precise break frequencies).
%
This spectral behaviour could be related to the approach to a possible break in electron energy distribution (as averaged for the whole SNR extent). 
It is possibly associated with cooling processes and/or intrinsic particle distribution features at the shock.

\begin{figure*}
  \centering
  \subfloat[IC443]{\includegraphics[width=0.5\textwidth]{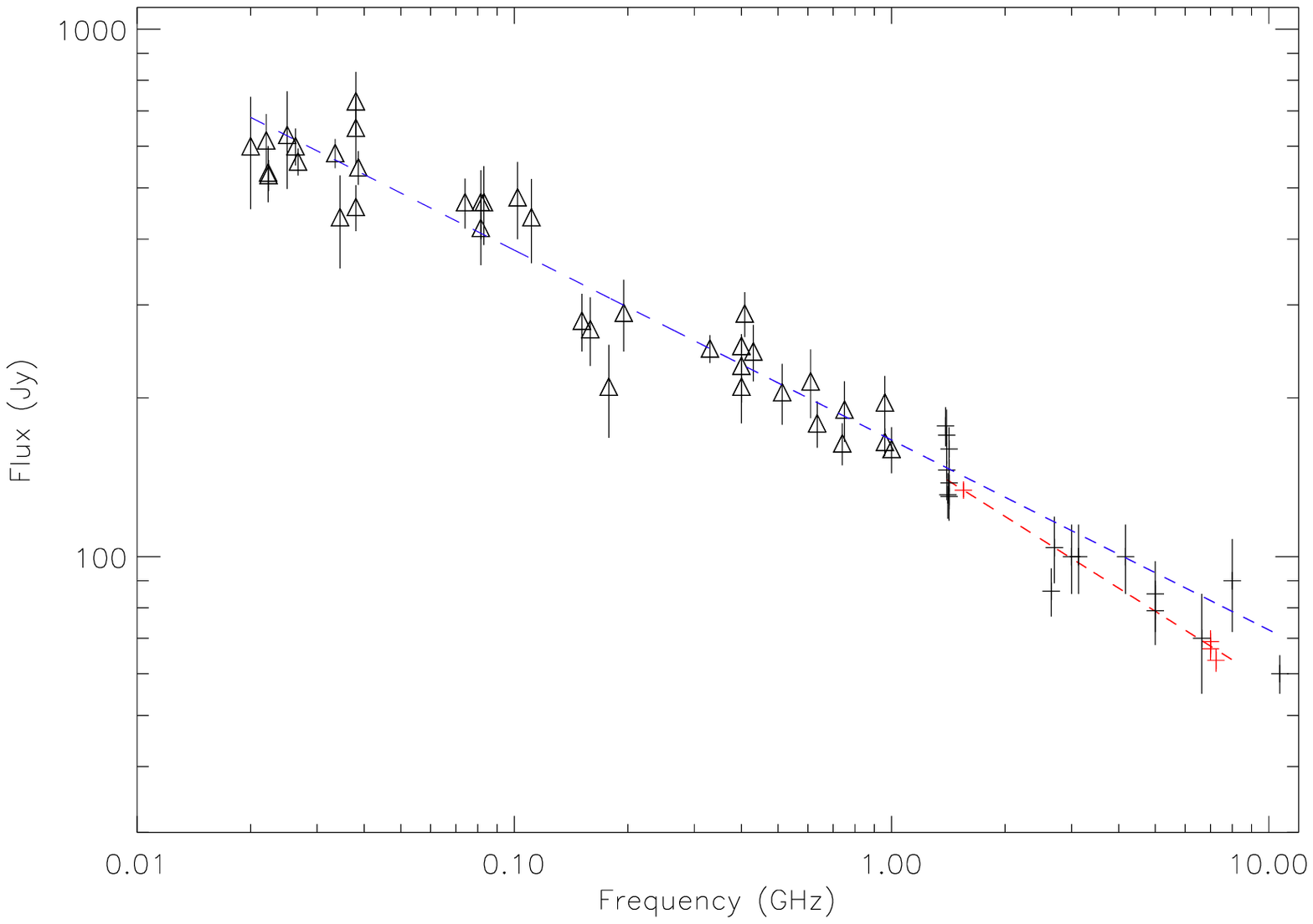}\label{fig:f1}}
  \hfill
  \subfloat[W44]{\includegraphics[width=0.5\textwidth]{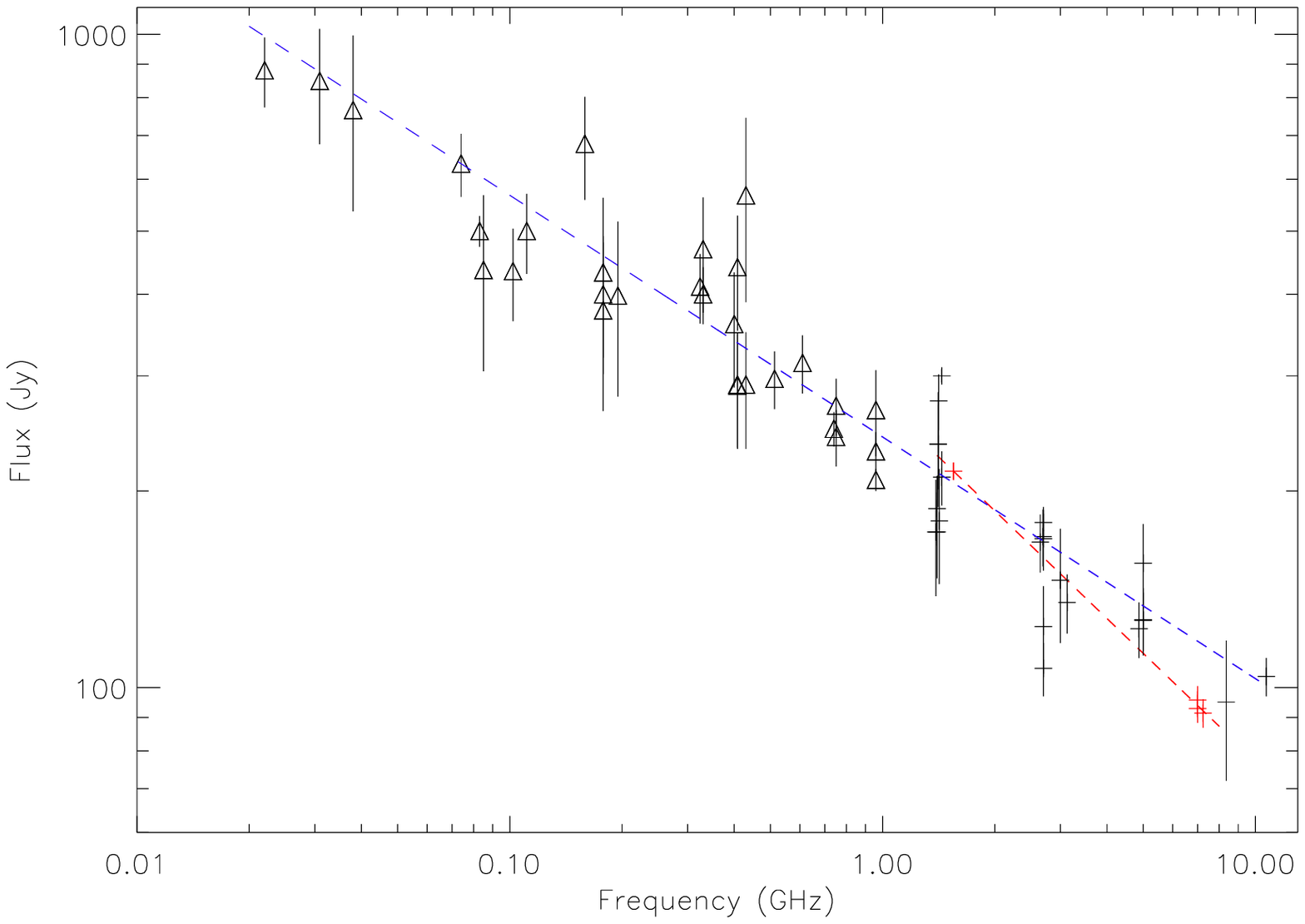}\label{fig:f2}}
  \caption{Integrated radio continuum spectra of IC443 (a) and W44 (b) obtained from the flux density values (black points) listed in \citet{Castelletti_11, Castelletti_07} for the frequency range $0.02-10.7$ GHz together with the fit provided by these authors (blue dashed line).
The spectral slopes derived from SRT data (Tables 2 and 3; red points and dashed line) at high frequencies ($1.5-7.2$ GHz) suggest a possible spectral steepening with respect to the lower frequency measurements ($0.02-1$ GHz; triangles).
}
\end{figure*}



A typical primary particle spectrum (both hadronic and leptonic) is expected to have a high energy cut-off which depends on the SNR age and other physical parameters (i.e. ambient density and magnetic field).
In fact, gamma-ray observations pointed out a steepening of the primary particle spectrum at energies of $\sim\,10$ GeV for W44 and $\sim\,100$ GeV for IC443 \citep{Cardillo_14,Ackermann_13,Giuliani_11B} that implies synchrotron break/cut-off
at frequencies $>$10 GHz . 
However, a secondary electron population produced by hadronic interactions could represent a major fraction of the whole leptonic plasma present in the SNR \citep{Cardillo_16,Lee_15}. 
These secondary hadronic electrons are expected to take $\sim 10\%$ of the primary particle energy.
Thus, a corresponding synchrotron spectrum change due to this particle population could be expected in the GHz range, in addition to the primary particle spectrum change at higher frequencies.

In order to properly investigate the actual parameters of region-dependent electron distributions, spatially-resolved,
high-frequency spectra are required.

\subsection{Disentangling different electron populations in IC443 and W44}

We clearly observe a significant spectral scatter along different SNR regions in our data.
The simplistic assumption of a single-electron population approximation typically postulated in SNR
physical modelling is thus manifestly hindered.
%
For both IC443 and W44, SRT images demonstrate that flat spectra correlate with the brightest SNR regions near the limbs or filaments, while the fainter central regions and halos display steeper spectra. Spectral flattening corresponding to bright regions of IC443 (e.g. toward the eastern boundary) was also evident from the analysis of VLA data at lower frequencies of $74-330$ MHz \citep{Castelletti_11}, although in this case this effect was related to thermal absorption, unlikely working at higher frequencies.

The significant spread in spectral index distributions within different SNR regions observed by SRT could in principle be related
to several and possibly concurring processes.

Thermal absorption (free-free) was invoked to explain (A) the low-frequency cut-offs ($<$50 MHz) observed in the integrated SNR spectrum of IC443, and (B) the apparent spread in spectral index distributions seen by the VLA at $<$1 GHz that is  possibly correlated with the nonuniform optical depth along the SNR.
The reported average free-free continuum optical depth derived for IC443 by \citet{Castelletti_11} is negligible for frequencies $>$ 50 MHz, but in principle absorption processes could be effective at higher frequencies if strong local enhancements of the optical depth are present. In fact, extrapolating the reported free-free optical depth local peaks for IC443 ($\tau_{74}\sim0.3$ at 74 MHz) to $\nu\,>\,$1 GHz, we obtain absorption coefficients exp(-$\tau_{74}(\nu/74_{\mathrm{ MHz}})^{-2.1}$)$\sim$1.
Thus, we can exclude that free-free thermal absorption could be responsible for the observed spectral index, region-dependent scatter at high frequencies.

The observed spread in the spectral index distribution of both IC443 and W44 could instead be related to an intrinsic variety in the primary and secondary electron spectra (spectral slopes and breaks)
 produced in shocks that are located in different SNR/PWN environments, -i.e.
several region-dependent electron populations are present.
Naive shock acceleration theory predicts a single particle spectrum that is a power law in momentum with spectral index $\sim$2, with
little fluctuation depending on the actual shock parameters  \citep[e.g.][and reference therein]{Sturner_97}. The resulting synchrotron spectral 
slope is then expected to be $\sim$0.5, a value that is compatible with our integrated spectra above $\sim$1 GHz. On the other hand, standard models fail to predict flat or inverted spectra as seen in bright SNR selected regions. Even in the ultrarelativistic regime and assuming a high shock compression factor ($\sim$4), electron spectra 
slopes are expected to be above $\sim$1.5 \citep{Ellison_96,Ellison_95}, thus
implying a synchrotron spectral index $>$0.2 that is incompatible with our results.

In order to overtake this incongruency, we note that
a spectral flattening effect related to the contribution of secondary hadronic electrons could be present in addition to the
primary electron canonical distribution \citep{Cardillo_16}. 
In fact, the correlation among bright flat-spectrum SNR regions and gamma-ray emission could represent a possible signature of enhanced hadronic
emission (and then significant secondary hadronic electrons injection).
For W44, gamma-ray emission seems to be associated with the bright radio
rims and filaments of the SNR \citep{Cardillo_14,Giuliani_11B,Abdo_10}.
Instead, the IC443 gamma-ray emission seems to be anticorrelated with SNR limbs, although it is still associated with a relatively bright radio filament (with a flatter spectrum 
w.r.t. average central region spectra) close to a molecular cloud enhancing IC/bremsstrahlung emission \citep{Humensky_15}.
Thus, the observed spread in spectral slopes could in principle reflect a region-dependent amount of secondary electron production, however this hypothesis cannot be clearly proven by present multi-wavelength data.

On the other hand, spectral steepening could also be related to strongly-enhanced,
region-dependent
cooling. However, this hypothesis is disfavoured by the evidence for a correlation among bright flat-spectrum radio regions and gamma-ray emission (i.e. the electron cooling to gamma-ray energies does not significantly affect radio spectra).
Furthermore, models of temporal evolution of nonthermal particle and photon spectra at different stages of shell-type SNR lifetime indicate that
no significant steepening  of the spectral index
due to synchrotron cooling
 is expected from a particle gas drifting away from the shock region
on a time-scale of $10^4 -10^5$ yrs \citep{Sturner_97}, unless a significant
local enhancement of the magnetic field can be envisaged. 
However, both primary and secondary electron distribution cut-offs could approach the synchrotron radio-emitting range on
these time-scales, thus providing a change in the synchrotron spectral index at high frequencies. In fact, as reported in section 5.2, a slight steepening of the integrated spectra of both IC443 and W44 at frequencies $>$1 GHz is speculated. Region-dependent spectral slopes could reflect the presence of different electron distribution cut-off energies.
In order to assess spectral curvature and cut-offs, high-resolution spectral imaging at frequencies $>$10 GHz is required.

\section{Conclusions}

In the frameworks of the SRT astronomical validation and the early science program,
we obtained single-dish, high-resolution maps of SNR IC443 and W44 at 7 GHz,
which provide accurate continuum flux density measurements.
By coupling them with SRT 1.5 GHz maps, we obtained spatially-resolved spectral measurements that are highlighting a spread in spectral slope distribution, ranging
from flat or slightly inverted spectra (up to $\alpha\sim-0.5$ for IC443) corresponding to bright radio structures,
 to relatively steep spectra ($\alpha$$\sim$0.7) in fainter radio regions of the SNRs.

We exclude that the observed region-dependent wide spread in spectral slope distribution could be related to absorption processes. Our high-frequency results can be directly related to distinct
electron populations in the SNRs including secondary hadronic electrons and resulting from different shocks conditions and/or undergoing
different cooling processes.

Integrated fluxes associated with the whole SNRs obtained by SRT in comparison with previous results in the literature support the evidence for a slight spectral steepening above $\sim$1 GHz for both sources, which could be related to primary electrons or more likely secondary hadronic electrons cut-offs.

Disentanglement among different theoretical possibilities for explaining the above findings could be provided through the analysis of further high-frequency/high-resolution imaging data. 
In particular, spatially-resolved, spectral curvature/break measurements 
could be obtained by coupling $1.5-7$ GHz maps with higher-frequency observations.

\section*{Acknowledgements}

The Sardinia Radio Telescope is funded by the Department of University and Research (MIUR), the Italian Space Agency (ASI), and the Autonomous Region of Sardinia (RAS), and is operated as a National Facility by the National Institute for Astrophysics (INAF). The development of the SARDARA backend has been funded by the Autonomous Region of Sardinia (RAS) using resources from the Regional Law 7/2007 "Promotion of the scientific research and technological innovation in Sardinia" in the context of the research project CRP-49231 (year 2011, PI A. Possenti): "High resolution sampling of the Universe in the radio band: an unprecedented instrument to understand the fundamental laws of the nature". 
S. Loru grateflully acknowledge the University of Cagliari and INAF for 
the financial support of her PhD scholarship.
F. Loi gratefully acknowledges Sardinia Regional Government for the financial support of her PhD scholarship (P.O.R. Sardegna F.S.E.
Operational Programme of the Autonomous Region of Sardinia, European Social Fund 2007-2013 - Axis IV Human Resources, Objective l.3, Line of
Activity l.3.1.).
M. Pilia was supported by the Sardinia Regional Government through the
project "Development of a Software Tool for the Study of Pulsars from Radio
to Gamma-rays using Multi-mission Data" (CRP-25476).




\bibliographystyle{mnras}
\bibliography{biblio} 

\begin{thebibliography}{}
\makeatletter
\relax
\def\mn@urlcharsother{\let\do\@makeother \do\$\do\&\do\#\do\^\do\_\do\%\do\~}
\def\mn@doi{\begingroup\mn@urlcharsother \@ifnextchar [ {\mn@doi@}
  {\mn@doi@[]}}
\def\mn@doi@[#1]#2{\def\@tempa{#1}\ifx\@tempa\@empty \href
  {http://dx.doi.org/#2} {doi:#2}\else \href {http://dx.doi.org/#2} {#1}\fi
  \endgroup}
\def\mn@eprint#1#2{\mn@eprint@#1:#2::\@nil}
\def\mn@eprint@arXiv#1{\href {http://arxiv.org/abs/#1} {{\tt arXiv:#1}}}
\def\mn@eprint@dblp#1{\href {http://dblp.uni-trier.de/rec/bibtex/#1.xml}
  {dblp:#1}}
\def\mn@eprint@#1:#2:#3:#4\@nil{\def\@tempa {#1}\def\@tempb {#2}\def\@tempc
  {#3}\ifx \@tempc \@empty \let \@tempc \@tempb \let \@tempb \@tempa \fi \ifx
  \@tempb \@empty \def\@tempb {arXiv}\fi \@ifundefined
  {mn@eprint@\@tempb}{\@tempb:\@tempc}{\expandafter \expandafter \csname
  mn@eprint@\@tempb\endcsname \expandafter{\@tempc}}}

\bibitem[\protect\citeauthoryear{{Abdo} et~al.,}{{Abdo} et~al.}{2010}]{Abdo_10}
{Abdo} A.~A.,  et~al., 2010, \mn@doi [Science] {10.1126/science.1182787}, \href
  {http://cdsads.u-strasbg.fr/abs/2010Sci...327.1103A} {327, 1103}

\bibitem[\protect\citeauthoryear{{Acero} et~al.,}{{Acero}
  et~al.}{2016}]{Acero_16}
{Acero} F.,  et~al., 2016, \mn@doi [\apjs] {10.3847/0067-0049/224/1/8}, \href
  {http://cdsads.u-strasbg.fr/abs/2016ApJS..224....8A} {224, 8}

\bibitem[\protect\citeauthoryear{{Ackermann} et~al.,}{{Ackermann}
  et~al.}{2013}]{Ackermann_13}
{Ackermann} M.,  et~al., 2013, \mn@doi [Science] {10.1126/science.1231160},
  \href {http://cdsads.u-strasbg.fr/abs/2013Sci...339..807A} {339, 807}

\bibitem[\protect\citeauthoryear{{Altenhoff}, {Downes}, {Goad}, {Maxwell}  \&
  {Rinehart}}{{Altenhoff} et~al.}{1970}]{Altenhoff_70}
{Altenhoff} W.~J.,  {Downes} D.,  {Goad} L.,  {Maxwell} A.,   {Rinehart} R.,
  1970, \aaps, \href {http://cdsads.u-strasbg.fr/abs/1970A%26AS....1..319A} {1,
  319}

\bibitem[\protect\citeauthoryear{{Altenhoff}, {Downes}, {Pauls}  \&
  {Schraml}}{{Altenhoff} et~al.}{1979}]{Altenhoff_79}
{Altenhoff} W.~J.,  {Downes} D.,  {Pauls} T.,   {Schraml} J.,  1979, \aaps,
  \href {http://cdsads.u-strasbg.fr/abs/1979A%26AS...35...23A} {35, 23}

\bibitem[\protect\citeauthoryear{{Amato}}{{Amato}}{2014}]{Amato_14}
{Amato} E.,  2014, \mn@doi [International Journal of Modern Physics D]
  {10.1142/S0218271814300134}, \href
  {http://cdsads.u-strasbg.fr/abs/2014IJMPD..2330013A} {23, 1430013}

\bibitem[\protect\citeauthoryear{{Baars}, {Genzel}, {Pauliny-Toth}  \&
  {Witzel}}{{Baars} et~al.}{1977}]{Baars_77}
{Baars} J.~W.~M.,  {Genzel} R.,  {Pauliny-Toth} I.~I.~K.,   {Witzel} A.,  1977,
  \aap, \href {http://cdsads.u-strasbg.fr/abs/1977A%26A....61...99B} {61, 99}

\bibitem[\protect\citeauthoryear{{Beard} \& {Kerr}}{{Beard} \&
  {Kerr}}{1969}]{Beard_69}
{Beard} M.,  {Kerr} F.~J.,  1969, Australian Journal of Physics, \href
  {http://cdsads.u-strasbg.fr/abs/1969AuJPh..22..121B} {22, 121}

\bibitem[\protect\citeauthoryear{{Blasi}}{{Blasi}}{2013}]{Blasi_13}
{Blasi} P.,  2013, \mn@doi [\aapr] {10.1007/s00159-013-0070-7}, \href
  {http://cdsads.u-strasbg.fr/abs/2013A%26ARv..21...70B} {21, 70}

\bibitem[\protect\citeauthoryear{{Bolli} et~al.,}{{Bolli}
  et~al.}{2015}]{Bolli_15}
{Bolli} P.,  et~al., 2015, \mn@doi [Journal of Astronomical Instrumentation]
  {10.1142/S2251171715500087}, \href
  {http://cdsads.u-strasbg.fr/abs/2015JAI.....450008B} {4, 1550008}

\bibitem[\protect\citeauthoryear{{Braun} \& {Strom}}{{Braun} \&
  {Strom}}{1986}]{Braun_86}
{Braun} R.,  {Strom} R.~G.,  1986, \aap, \href
  {http://cdsads.u-strasbg.fr/abs/1986A%26A...164..193B} {164, 193}

\bibitem[\protect\citeauthoryear{{Burton}, {Geballe}, {Brand}  \&
  {Webster}}{{Burton} et~al.}{1988}]{Burton_88}
{Burton} M.~G.,  {Geballe} T.~R.,  {Brand} P.~W.~J.~L.,   {Webster} A.~S.,
  1988, \mn@doi [\mnras] {10.1093/mnras/231.3.617}, \href
  {http://cdsads.u-strasbg.fr/abs/1988MNRAS.231..617B} {231, 617}

\bibitem[\protect\citeauthoryear{{Buttu} et~al.,}{{Buttu}
  et~al.}{2012}]{Buttu_12}
{Buttu} M.,  et~al., 2012, in Software and Cyberinfrastructure for Astronomy
  II. p. 84512L, \mn@doi{10.1117/12.925387}

\bibitem[\protect\citeauthoryear{{Cardillo} et~al.,}{{Cardillo}
  et~al.}{2014}]{Cardillo_14}
{Cardillo} M.,  et~al., 2014, \mn@doi [\aap] {10.1051/0004-6361/201322685},
  \href {http://cdsads.u-strasbg.fr/abs/2014A%26A...565A..74C} {565, A74}

\bibitem[\protect\citeauthoryear{{Cardillo}, {Amato}  \& {Blasi}}{{Cardillo}
  et~al.}{2016}]{Cardillo_16}
{Cardillo} M.,  {Amato} E.,   {Blasi} P.,  2016, preprint, \href
  {http://cdsads.u-strasbg.fr/abs/2016arXiv160402321C} {} (\mn@eprint {arXiv}
  {1604.02321})

\bibitem[\protect\citeauthoryear{{Castelletti}, {Dubner}, {Brogan}  \&
  {Kassim}}{{Castelletti} et~al.}{2007}]{Castelletti_07}
{Castelletti} G.,  {Dubner} G.,  {Brogan} C.,   {Kassim} N.~E.,  2007, \mn@doi
  [\aap] {10.1051/0004-6361:20077062}, \href
  {http://cdsads.u-strasbg.fr/abs/2007A%26A...471..537C} {471, 537}

\bibitem[\protect\citeauthoryear{{Castelletti}, {Dubner}, {Clarke}  \&
  {Kassim}}{{Castelletti} et~al.}{2011}]{Castelletti_11}
{Castelletti} G.,  {Dubner} G.,  {Clarke} T.,   {Kassim} N.~E.,  2011, \mn@doi
  [\aap] {10.1051/0004-6361/201016081}, \href
  {http://cdsads.u-strasbg.fr/abs/2011A%26A...534A..21C} {534, A21}

\bibitem[\protect\citeauthoryear{{Cesarsky}, {Cox}, {Pineau des For{\^e}ts},
  {van Dishoeck}, {Boulanger}  \& {Wright}}{{Cesarsky}
  et~al.}{1999}]{Cesarsky_99}
{Cesarsky} D.,  {Cox} P.,  {Pineau des For{\^e}ts} G.,  {van Dishoeck} E.~F.,
  {Boulanger} F.,   {Wright} C.~M.,  1999, \aap, \href
  {http://cdsads.u-strasbg.fr/abs/1999A%26A...348..945C} {348, 945}

\bibitem[\protect\citeauthoryear{{Chevalier}}{{Chevalier}}{1999}]{Chevalier_99}
{Chevalier} R.~A.,  1999, \mn@doi [\apj] {10.1086/306710}, \href
  {http://cdsads.u-strasbg.fr/abs/1999ApJ...511..798C} {511, 798}

\bibitem[\protect\citeauthoryear{{Clark} \& {Caswell}}{{Clark} \&
  {Caswell}}{1976}]{Clark_76}
{Clark} D.~H.,  {Caswell} J.~L.,  1976, \mn@doi [\mnras]
  {10.1093/mnras/174.2.267}, \href
  {http://cdsads.u-strasbg.fr/abs/1976MNRAS.174..267C} {174, 267}

\bibitem[\protect\citeauthoryear{{Dickel}}{{Dickel}}{1971}]{Dickel_71}
{Dickel} J.~R.,  1971, \mn@doi [\pasp] {10.1086/129133}, \href
  {http://cdsads.u-strasbg.fr/abs/1971PASP...83..343D} {83, 343}

\bibitem[\protect\citeauthoryear{{Dickman}, {Snell}, {Ziurys}  \&
  {Huang}}{{Dickman} et~al.}{1992}]{Dickman_92}
{Dickman} R.~L.,  {Snell} R.~L.,  {Ziurys} L.~M.,   {Huang} Y.-L.,  1992,
  \mn@doi [\apj] {10.1086/171987}, \href
  {http://cdsads.u-strasbg.fr/abs/1992ApJ...400..203D} {400, 203}

\bibitem[\protect\citeauthoryear{{Duin} \& {van der Laan}}{{Duin} \& {van der
  Laan}}{1975}]{Duin_75}
{Duin} R.~M.,  {van der Laan} H.,  1975, \aap, \href
  {http://cdsads.u-strasbg.fr/abs/1975A%26A....40..111D} {40, 111}

\bibitem[\protect\citeauthoryear{{Egron}, {Pellizzoni}, {Iacolina}, {Loru},
  {Righini}, {Trois}  \& {SRT Astrophysical Validation Team}}{{Egron}
  et~al.}{2016}]{Egron_16}
{Egron} E.,  {Pellizzoni} A.,  {Iacolina} M.~N.,  {Loru} S.,  {Righini} S.,
  {Trois} A.,   {SRT Astrophysical Validation Team} 2016, INAF - Osservatorio
  Astronomico di Cagliari. Internal Report N.59, \href
  {http://www.oa-cagliari.inaf.it/area.php?page_id=10} {}

\bibitem[\protect\citeauthoryear{{Ellison}, {Baring}  \& {Jones}}{{Ellison}
  et~al.}{1995}]{Ellison_95}
{Ellison} D.~C.,  {Baring} M.~G.,   {Jones} F.~C.,  1995, \mn@doi [\apj]
  {10.1086/176447}, \href {http://cdsads.u-strasbg.fr/abs/1995ApJ...453..873E}
  {453, 873}

\bibitem[\protect\citeauthoryear{{Ellison}, {Baring}  \& {Jones}}{{Ellison}
  et~al.}{1996}]{Ellison_96}
{Ellison} D.~C.,  {Baring} M.~G.,   {Jones} F.~C.,  1996, \mn@doi [\apj]
  {10.1086/178213}, \href {http://cdsads.u-strasbg.fr/abs/1996ApJ...473.1029E}
  {473, 1029}

\bibitem[\protect\citeauthoryear{{Frail}, {Giacani}, {Goss}  \&
  {Dubner}}{{Frail} et~al.}{1996}]{Frail_96}
{Frail} D.~A.,  {Giacani} E.~B.,  {Goss} W.~M.,   {Dubner} G.,  1996, \mn@doi
  [\apjl] {10.1086/310103}, \href
  {http://cdsads.u-strasbg.fr/abs/1996ApJ...464L.165F} {464, L165}

\bibitem[\protect\citeauthoryear{{Gao}, {Han}, {Reich}, {Reich}, {Sun}  \&
  {Xiao}}{{Gao} et~al.}{2011}]{Gao_11}
{Gao} X.~Y.,  {Han} J.~L.,  {Reich} W.,  {Reich} P.,  {Sun} X.~H.,   {Xiao} L.,
   2011, \mn@doi [\aap] {10.1051/0004-6361/201016311}, \href
  {http://cdsads.u-strasbg.fr/abs/2011A%26A...529A.159G} {529, A159}

\bibitem[\protect\citeauthoryear{{G{\'e}nova-Santos}
  et~al.,}{{G{\'e}nova-Santos} et~al.}{2016}]{Genova_16}
{G{\'e}nova-Santos} R.,  et~al., 2016, \mn@doi [\mnras]
  {10.1093/mnras/stw2503}, \href
  {http://cdsads.u-strasbg.fr/abs/2016MNRAS.tmp.1497G} {}

\bibitem[\protect\citeauthoryear{{Giacani}, {Dubner}, {Kassim}, {Frail},
  {Goss}, {Winkler}  \& {Williams}}{{Giacani} et~al.}{1997}]{Giacani_97}
{Giacani} E.~B.,  {Dubner} G.~M.,  {Kassim} N.~E.,  {Frail} D.~A.,  {Goss}
  W.~M.,  {Winkler} P.~F.,   {Williams} B.~F.,  1997, \mn@doi [\aj]
  {10.1086/118352}, \href {http://cdsads.u-strasbg.fr/abs/1997AJ....113.1379G}
  {113, 1379}

\bibitem[\protect\citeauthoryear{{Ginzburg} \& {Syrovatskii}}{{Ginzburg} \&
  {Syrovatskii}}{1964}]{Ginzburg_64}
{Ginzburg} V.~L.,  {Syrovatskii} S.~I.,  1964, {The Origin of Cosmic Rays}

\bibitem[\protect\citeauthoryear{{Giuliani} \& {AGILE Team}}{{Giuliani} \&
  {AGILE Team}}{2011}]{Giuliani_11A}
{Giuliani} G.,  {AGILE Team} 2011, \memsai, \href
  {http://cdsads.u-strasbg.fr/abs/2011MmSAI..82..747G} {82, 747}

\bibitem[\protect\citeauthoryear{{Giuliani} et~al.,}{{Giuliani}
  et~al.}{2011}]{Giuliani_11B}
{Giuliani} A.,  et~al., 2011, \mn@doi [\apjl] {10.1088/2041-8205/742/2/L30},
  \href {http://cdsads.u-strasbg.fr/abs/2011ApJ...742L..30G} {742, L30}

\bibitem[\protect\citeauthoryear{{Green}}{{Green}}{1986}]{Green_86}
{Green} D.~A.,  1986, \mn@doi [\mnras] {10.1093/mnras/221.2.473}, \href
  {http://cdsads.u-strasbg.fr/abs/1986MNRAS.221..473G} {221, 473}

\bibitem[\protect\citeauthoryear{{Green}}{{Green}}{2014}]{Green_14}
{Green} D.~A.,  2014, Bulletin of the Astronomical Society of India, \href
  {http://cdsads.u-strasbg.fr/abs/2014BASI...42...47G} {42, 47}

\bibitem[\protect\citeauthoryear{{H.E.S.S.~Collaboration}
  et~al.,}{{H.E.S.S.~Collaboration} et~al.}{2011}]{Hess_11}
{H.E.S.S.~Collaboration} et~al., 2011, \mn@doi [\aap]
  {10.1051/0004-6361/201016425}, \href
  {http://cdsads.u-strasbg.fr/abs/2011A%26A...531A..81H} {531, A81}

\bibitem[\protect\citeauthoryear{{Hollinger} \& {Hobbs}}{{Hollinger} \&
  {Hobbs}}{1966}]{Hollinger_66}
{Hollinger} J.~P.,  {Hobbs} R.~W.,  1966, \mn@doi [Science]
  {10.1126/science.153.3744.1633}, \href
  {http://cdsads.u-strasbg.fr/abs/1966Sci...153.1633H} {153, 1633}

\bibitem[\protect\citeauthoryear{{Humensky} \& {the VERITAS
  Collaboration}}{{Humensky} \& {the VERITAS
  Collaboration}}{2015}]{Humensky_15}
{Humensky} B.,  {the VERITAS Collaboration} 2015, preprint, \href
  {http://cdsads.u-strasbg.fr/abs/2015arXiv151201911H} {} (\mn@eprint {arXiv}
  {1512.01911})

\bibitem[\protect\citeauthoryear{{Jones}, {Smith}  \& {Angelini}}{{Jones}
  et~al.}{1993}]{Jones_93}
{Jones} L.~R.,  {Smith} A.,   {Angelini} L.,  1993, \mn@doi [\mnras]
  {10.1093/mnras/265.3.631}, \href
  {http://cdsads.u-strasbg.fr/abs/1993MNRAS.265..631J} {265, 631}

\bibitem[\protect\citeauthoryear{{Lee}, {Koo}, {Yun}, {Stanimirovi{\'c}},
  {Heiles}  \& {Heyer}}{{Lee} et~al.}{2008}]{Lee_08}
{Lee} J.-J.,  {Koo} B.-C.,  {Yun} M.~S.,  {Stanimirovi{\'c}} S.,  {Heiles} C.,
   {Heyer} M.,  2008, \mn@doi [\aj] {10.1088/0004-6256/135/3/796}, \href
  {http://cdsads.u-strasbg.fr/abs/2008AJ....135..796L} {135, 796}

\bibitem[\protect\citeauthoryear{{Lee}, {Patnaude}, {Raymond}, {Nagataki},
  {Slane}  \& {Ellison}}{{Lee} et~al.}{2015}]{Lee_15}
{Lee} S.-H.,  {Patnaude} D.~J.,  {Raymond} J.~C.,  {Nagataki} S.,  {Slane}
  P.~O.,   {Ellison} D.~C.,  2015, \mn@doi [\apj] {10.1088/0004-637X/806/1/71},
  \href {http://cdsads.u-strasbg.fr/abs/2015ApJ...806...71L} {806, 71}

\bibitem[\protect\citeauthoryear{{Leslie}}{{Leslie}}{1960}]{Leslie_60}
{Leslie} P.~R.~R.,  1960, The Observatory, \href
  {http://cdsads.u-strasbg.fr/abs/1960Obs....80...23L} {80, 23}

\bibitem[\protect\citeauthoryear{{Mavromatakis}, {Boumis}  \&
  {Goudis}}{{Mavromatakis} et~al.}{2003}]{Mavromatakis_03}
{Mavromatakis} F.,  {Boumis} P.,   {Goudis} C.~D.,  2003, \mn@doi [\aap]
  {10.1051/0004-6361:20030603}, \href
  {http://cdsads.u-strasbg.fr/abs/2003A%26A...405..591M} {405, 591}

\bibitem[\protect\citeauthoryear{{Olbert}, {Clearfield}, {Williams}, {Keohane}
  \& {Frail}}{{Olbert} et~al.}{2001}]{Olbert_01}
{Olbert} C.~M.,  {Clearfield} C.~R.,  {Williams} N.~E.,  {Keohane} J.~W.,
  {Frail} D.~A.,  2001, \mn@doi [\apjl] {10.1086/321708}, \href
  {http://cdsads.u-strasbg.fr/abs/2001ApJ...554L.205O} {554, L205}

\bibitem[\protect\citeauthoryear{{Oni{\'c}}}{{Oni{\'c}}}{2015}]{Onic_15}
{Oni{\'c}} D.,  2015, \mn@doi [Serbian Astronomical Journal]
  {10.2298/SAJ150715004O}, \href
  {http://cdsads.u-strasbg.fr/abs/2015SerAJ.191...29O} {191, 29}

\bibitem[\protect\citeauthoryear{{Orfei}, {Morsiani}, {Zacchiroli},
  {Maccaferri}, {Roda}  \& {Fiocchi}}{{Orfei} et~al.}{2004}]{Orfei_04}
{Orfei} A.,  {Morsiani} M.,  {Zacchiroli} G.,  {Maccaferri} G.,  {Roda} J.,
  {Fiocchi} F.,  2004, in {Antebi} J.,  {Lemke} D.,  eds,  \procspie Vol. 5495,
  Astronomical Structures and Mechanisms Technology. pp 116--125,
  \mn@doi{10.1117/12.548944}

\bibitem[\protect\citeauthoryear{{Paladini}, {Burigana}, {Davies}, {Maino},
  {Bersanelli}, {Cappellini}, {Platania}  \& {Smoot}}{{Paladini}
  et~al.}{2003}]{Paladini_03}
{Paladini} R.,  {Burigana} C.,  {Davies} R.~D.,  {Maino} D.,  {Bersanelli} M.,
  {Cappellini} B.,  {Platania} P.,   {Smoot} G.,  2003, \mn@doi [\aap]
  {10.1051/0004-6361:20021466}, \href
  {http://cdsads.u-strasbg.fr/abs/2003A%26A...397..213P} {397, 213}

\bibitem[\protect\citeauthoryear{{Pauliny-Toth}, {Wade}  \&
  {Heeschen}}{{Pauliny-Toth} et~al.}{1966}]{Pauliny-Toth_66}
{Pauliny-Toth} I.~I.~K.,  {Wade} C.~M.,   {Heeschen} D.~S.,  1966, \mn@doi
  [\apjs] {10.1086/190137}, \href
  {http://cdsads.u-strasbg.fr/abs/1966ApJS...13...65P} {13, 65}

\bibitem[\protect\citeauthoryear{{Perley} \& {Butler}}{{Perley} \&
  {Butler}}{2013}]{Perley_13}
{Perley} R.~A.,  {Butler} B.~J.,  2013, \mn@doi [\apjs]
  {10.1088/0067-0049/204/2/19}, \href
  {http://cdsads.u-strasbg.fr/abs/2013ApJS..204...19P} {204, 19}

\bibitem[\protect\citeauthoryear{{Petre}, {Szymkowiak}, {Seward}  \&
  {Willingale}}{{Petre} et~al.}{1988}]{Petre_88}
{Petre} R.,  {Szymkowiak} A.~E.,  {Seward} F.~D.,   {Willingale} R.,  1988,
  \mn@doi [\apj] {10.1086/166922}, \href
  {http://cdsads.u-strasbg.fr/abs/1988ApJ...335..215P} {335, 215}

\bibitem[\protect\citeauthoryear{{Petre}, {Kuntz}  \& {Shelton}}{{Petre}
  et~al.}{2002}]{Petre_02}
{Petre} R.,  {Kuntz} K.~D.,   {Shelton} R.~L.,  2002, \mn@doi [\apj]
  {10.1086/342672}, \href {http://cdsads.u-strasbg.fr/abs/2002ApJ...579..404P}
  {579, 404}

\bibitem[\protect\citeauthoryear{{Reach}, {Rho}  \& {Jarrett}}{{Reach}
  et~al.}{2005}]{Reach_05}
{Reach} W.~T.,  {Rho} J.,   {Jarrett} T.~H.,  2005, \mn@doi [\apj]
  {10.1086/425855}, \href {http://cdsads.u-strasbg.fr/abs/2005ApJ...618..297R}
  {618, 297}

\bibitem[\protect\citeauthoryear{{Reich}, {Zhang}  \& {F{\"u}rst}}{{Reich}
  et~al.}{2003}]{Reich_03}
{Reich} W.,  {Zhang} X.,   {F{\"u}rst} E.,  2003, \mn@doi [\aap]
  {10.1051/0004-6361:20030939}, \href
  {http://cdsads.u-strasbg.fr/abs/2003A%26A...408..961R} {408, 961}

\bibitem[\protect\citeauthoryear{{Reynolds}}{{Reynolds}}{2008}]{Reynolds_08}
{Reynolds} S.~P.,  2008, \mn@doi [\araa]
  {10.1146/annurev.astro.46.060407.145237}, \href
  {http://cdsads.u-strasbg.fr/abs/2008ARA%26A..46...89R} {46, 89}

\bibitem[\protect\citeauthoryear{{Rho} \& {Petre}}{{Rho} \&
  {Petre}}{1998}]{Rho_98}
{Rho} J.,  {Petre} R.,  1998, \mn@doi [\apjl] {10.1086/311538}, \href
  {http://cdsads.u-strasbg.fr/abs/1998ApJ...503L.167R} {503, L167}

\bibitem[\protect\citeauthoryear{{Scheuer}}{{Scheuer}}{1963}]{Scheuer_63}
{Scheuer} P.~A.~G.,  1963, The Observatory, \href
  {http://cdsads.u-strasbg.fr/abs/1963Obs....83...56S} {83, 56}

\bibitem[\protect\citeauthoryear{{Seta}, {Hasegawa}, {Sakamoto}, {Oka},
  {Sawada}, {Inutsuka}, {Koyama}  \& {Hayashi}}{{Seta} et~al.}{2004}]{Seta_04}
{Seta} M.,  {Hasegawa} T.,  {Sakamoto} S.,  {Oka} T.,  {Sawada} T.,  {Inutsuka}
  S.-i.,  {Koyama} H.,   {Hayashi} M.,  2004, \mn@doi [\aj] {10.1086/381058},
  \href {http://cdsads.u-strasbg.fr/abs/2004AJ....127.1098S} {127, 1098}

\bibitem[\protect\citeauthoryear{{Smith}, {Jones}, {Watson}, {Willingale},
  {Wood}  \& {Seward}}{{Smith} et~al.}{1985}]{Smith_85}
{Smith} A.,  {Jones} L.~R.,  {Watson} M.~G.,  {Willingale} R.,  {Wood} N.,
  {Seward} F.~D.,  1985, \mn@doi [\mnras] {10.1093/mnras/217.1.99}, \href
  {http://cdsads.u-strasbg.fr/abs/1985MNRAS.217...99S} {217, 99}

\bibitem[\protect\citeauthoryear{{Snell}, {Hollenbach}, {Howe}, {Neufeld},
  {Kaufman}, {Melnick}, {Bergin}  \& {Wang}}{{Snell} et~al.}{2005}]{Snell_05}
{Snell} R.~L.,  {Hollenbach} D.,  {Howe} J.~E.,  {Neufeld} D.~A.,  {Kaufman}
  M.~J.,  {Melnick} G.~J.,  {Bergin} E.~A.,   {Wang} Z.,  2005, \mn@doi [\apj]
  {10.1086/427231}, \href {http://cdsads.u-strasbg.fr/abs/2005ApJ...620..758S}
  {620, 758}

\bibitem[\protect\citeauthoryear{{Sturner}, {Skibo}, {Dermer}  \&
  {Mattox}}{{Sturner} et~al.}{1997}]{Sturner_97}
{Sturner} S.~J.,  {Skibo} J.~G.,  {Dermer} C.~D.,   {Mattox} J.~R.,  1997,
  \apj, \href {http://cdsads.u-strasbg.fr/abs/1997ApJ...490..619S} {490, 619}

\bibitem[\protect\citeauthoryear{{Sun}, {Reich}, {Reich}, {Xiao}, {Gao}  \&
  {Han}}{{Sun} et~al.}{2011}]{Sun_11}
{Sun} X.~H.,  {Reich} P.,  {Reich} W.,  {Xiao} L.,  {Gao} X.~Y.,   {Han} J.~L.,
   2011, \mn@doi [\aap] {10.1051/0004-6361/201117693}, \href
  {http://cdsads.u-strasbg.fr/abs/2011A%26A...536A..83S} {536, A83}

\bibitem[\protect\citeauthoryear{{Swartz} et~al.,}{{Swartz}
  et~al.}{2015}]{Swartz_15}
{Swartz} D.~A.,  et~al., 2015, \mn@doi [\apj] {10.1088/0004-637X/808/1/84},
  \href {http://cdsads.u-strasbg.fr/abs/2015ApJ...808...84S} {808, 84}

\bibitem[\protect\citeauthoryear{{Troja}, {Bocchino}, {Miceli}  \&
  {Reale}}{{Troja} et~al.}{2008}]{Troja_08}
{Troja} E.,  {Bocchino} F.,  {Miceli} M.,   {Reale} F.,  2008, \mn@doi [\aap]
  {10.1051/0004-6361:20079123}, \href
  {http://cdsads.u-strasbg.fr/abs/2008A%26A...485..777T} {485, 777}

\bibitem[\protect\citeauthoryear{{Valente} et~al.,}{{Valente}
  et~al.}{2010}]{Valente_10}
{Valente} G.,  et~al., 2010, in Millimeter, Submillimeter, and Far-Infrared
  Detectors and Instrumentation for Astronomy V. p. 774126,
  \mn@doi{10.1117/12.857306}

\bibitem[\protect\citeauthoryear{{Valente} et~al.,}{{Valente}
  et~al.}{2016}]{Valente_16}
{Valente} G.,  et~al., 2016, in Millimeter, Submillimeter, and Far-Infrared
  Detectors and Instrumentation for Astronomy VIII. p. 991425,
  \mn@doi{10.1117/12.2232880}

\bibitem[\protect\citeauthoryear{{Wolszczan}, {Cordes}  \& {Dewey}}{{Wolszczan}
  et~al.}{1991}]{Wolszczan_91}
{Wolszczan} A.,  {Cordes} J.~M.,   {Dewey} R.~J.,  1991, \mn@doi [\apjl]
  {10.1086/186033}, \href {http://cdsads.u-strasbg.fr/abs/1991ApJ...372L..99W}
  {372, L99}

\bibitem[\protect\citeauthoryear{{Wootten}}{{Wootten}}{1977}]{Wootten_77}
{Wootten} H.~A.,  1977, \mn@doi [\apj] {10.1086/155485}, \href
  {http://cdsads.u-strasbg.fr/abs/1977ApJ...216..440W} {216, 440}

\bibitem[\protect\citeauthoryear{{Yoshiike} et~al.,}{{Yoshiike}
  et~al.}{2013}]{Yoshiike_13}
{Yoshiike} S.,  et~al., 2013, \mn@doi [\apj] {10.1088/0004-637X/768/2/179},
  \href {http://cdsads.u-strasbg.fr/abs/2013ApJ...768..179Y} {768, 179}

\makeatother
\end{thebibliography}










\bsp	
\label{lastpage}
\end{document}